\documentstyle[sprocl,epsf,epsfig]{article}

\def\gappeq{\mathrel{\rlap{\raise.5ex\hbox{$>$}}{\lower.5ex\hbox{$\sim$}}}}
\def\lappeq{\mathrel{\rlap{\raise.5ex\hbox{$<$}}{\lower.5ex\hbox{$\sim$}}}}

\def\beq{\begin{equation}}
\def\eeq{\end{equation}}
\def\bea{\begin{eqnarray}}
\def\eea{\end{eqnarray}}
\def\bq{\begin{quote}}
\def\eq{\end{quote}}

\def\APP{{\it Acta Phys.Pol.} }

\def\IJMP{{\it Int.J.Mod.Phys.} }

\def\LNC{{\it Lett. Nuovo Cimento} }

\def\MPL{{\it Mod.Phys.Lett.} }

\def\NC{{\it Nuovo Cimento} }
\def\NP{{\it Nucl.Phys.} }

\def\PL{{\it Phys.Lett.} }

\def\PR{{\it Phys.Rev.} }
\def\PRL{{\it Phys.Rev.Lett.} }
\def\PRTS{{\it Physics Reports} }

\def\PTP{{\it Progr.Theor.Phys.} }

\def\ZP{{\it Z.Phys.} }

\def\gappeq{\mathrel{\rlap {\raise.5ex\hbox{$>$}}
{\lower.5ex\hbox{$\sim$}}}}

\def\lappeq{\mathrel{\rlap{\raise.5ex\hbox{$<$}}
{\lower.5ex\hbox{$\sim$}}}}

\begin{document}
\thispagestyle{empty}
\vspace*{-3cm}
\begin{flushright}
{CERN-TH/97-131}\\
\end{flushright}
\vspace*{5mm}
\begin{center}
{\bf PHENOMENOLOGY OF LEP 2 PHYSICS}\\
\vspace*{1cm} 
{\bf John ELLIS} \\
\vspace{0.3cm}
Theoretical Physics Division, CERN \\
CH - 1211 Geneva 23 \\
\vspace*{2cm}  
{\bf ABSTRACT} \\ 
\end{center}
\vspace*{5mm}
\noindent
Various aspects of physics at LEP 2 are reviewed from a phenomenological
point of view. We first discuss the search for Higgs bosons, which
might be relatively light if indications from precision electroweak data
and supersymmetry are correct. Then $WW$ physics is discussed, with 
particular emphasis on the problems in measuring $m_W$ by the kinematic 
reconstruction of events with purely hadronic final states. Finally, 
possible manifestations of supersymmetry are reviewed, comparing 
different scenarios for the lightest supersymmetric particle, which may 
or may not be stable. Possible scenarios for $R$ violation at HERA and 
future searches for supersymmetry at the LHC are also mentioned briefly.
\vspace*{2cm}

\begin{center}
{\it Lectures presented at the Lake Louise Winter Institute}\\
{\it Lake Louise, Alberta, Canada}\\
{\it February 1997} 
\end{center}
\noindent 

\begin{flushleft} 
CERN-TH/97-131 \\
July 1997
\end{flushleft}

\textwidth=14.5cm
\vfil\eject
\setcounter{page}{1}
\title{PHENOMENOLOGY OF LEP 2 PHYSICS}

\author{ John Ellis}
\address{
Theoretical Physics Division, CERN \\
CH - 1211 Geneva 23}

\maketitle\abstracts{Various aspects of physics at LEP 2 are reviewed from a
phenomenological point of view. We first discuss the search for Higgs bosons,
which might be relatively light if indications from the precision electroweak
data and supersymmetry are correct. Then $WW$ physics is discussed, with
particular emphasis on the problems in measuring $m_W$ by the kinematic
reconstruction of events with purely hadronic final states. Finally, possible
manifestations of supersymmetry are reviewed, comparing different scenarios for
the lightest supersymmetric particle, which may or may not be stable. Possible
scenarios for $R$ violation at HERA and future searches for supersymmetry at
the LHC are also mentioned briefly.}

\section*{Preamble}

The cross-sections for the most important physics processes that we expect (or
hope) to see at LEP 2 are shown in Fig. 1~\cite{LEP2}. The reactions
$e^+e^-\rightarrow \bar
ff$, which have been the bread and butter of LEP 1 physics, become backgrounds to
be fought at LEP 2. The reaction $e^+e^-\rightarrow W^+W^-$ is likely to provide
the new bread and butter for LEP 2. The reaction $e^+e^-\rightarrow Z^0Z^0$ has
not excited much interest so far, though it may provide a pesky background to
the Higgs search if $M_H \sim$ 90 GeV. We can but hope that the reaction
$e^+e^-\rightarrow Z + H$~\cite{ZH} is accessible to LEP 2!

\begin{figure}
\hglue2cm
\epsfig{figure=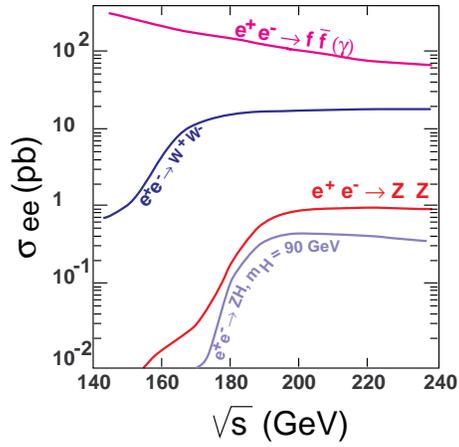,width=6cm}
\caption[]{
Cross-sections for the main processes to be measured at LEP~2.}
\end{figure}

\begin{figure}
\hglue3cm
\epsfig{figure=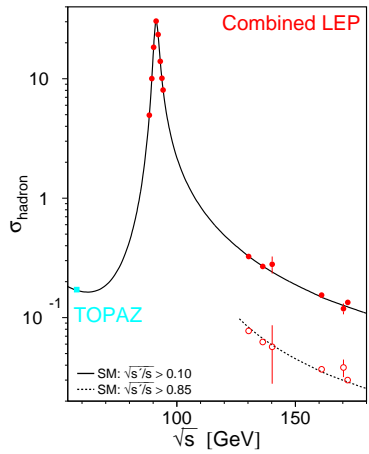,width=5cm}
\caption[]{
Measured $e^+e^-$ hadron cross-sections at LEP and from the lower-energy TOPAZ
experiment at KEK. At high energies, the measured cross-section depends strongly on
the cut on the observed energy $\sqrt{s^\prime}$ that is imposed.}
\end{figure}

The upgrades of LEP using superconducting RF cavities started in the
Autumn of 1995 with the first runs at $E_{cm}$ = 130 and
136 GeV, and a little data at 140
GeV, known collectively as LEP 1.5 \footnote{Since $E_{cm} \simeq 1.5 m_Z$.}.
These were followed in the Summer of 1996 by the first runs above the $W^+W^-$
threshold, at $E_{cm}$ = 161.3 GeV, known as LEP 2W. Later in 1996 came runs at
170/172 GeV referred to here as LEP 2. Cross-sections for $e^+e^-\rightarrow \bar
ff$ measured at LEP 1, 1.5, 2W and 2 energies are shown in Fig.
2~\cite{LEPEWWG}. Subsequent
plans include runs at $E_{cm} \simeq$ 184 GeV during 1997, and at $E_{cm} \simeq$
192 GeV in 1998 and 1999. There is a possibility of increasing the maximum LEP 2
energy to about 200 GeV if the superconducting RF can be coaxed to higher
accelerating fields $\sim$ 7 MeV/m with the aid of the LHC cryogenics system.
Running LEP in the year 2000 also seems to be compatible with the LHC
construction schedule, though this will cause earth motion that may complicate
LEP running from 1999 onwards. It remains to be seen whether physics will
warrant, and finances permit, running LEP during the year 2000. But enough of
this preamble: now is the time to address the first item on the physics agenda for
LEP 2.

\section{In pursuit of the Higgs boson}

\subsection{Precision Electroweak Measurements}

These are available from a large range of energies and distance scales,
extending from parity violation in atoms through various fixed-target lepton
scattering experiments to high-energy experiments at $e^+e^-, \bar pp$ and $ep$
colliders. Among these, the largest weight is currently carried by the $e^+e^-$
experiments at the $Z^0$ peak at LEP and the SLC, together with the Tevatron
collider experiments CDF and $D0$. The high-$Q^2$ events from
HERA~\cite{H1,ZEUS} are another story $\ldots$

The basic measurements on the $Z^0$ peak~\cite{Zpeak} include the total
hadronic cross-section, which is given at the tree level by
\beq
\sigma_h^0 = {12\pi\over m^2_Z}~~{\Gamma_{ee}\Gamma_{had}\over\Gamma^2_Z}
\label{oneone}
\eeq
which is reduced substantially by radiative corrections (principally
Initial-State Radiation, ISR) to  about 30 nb. The total $Z$ decay rate may be
decomposed as
\beq
\Gamma_Z = \Gamma_\ell + \Gamma_\mu + \Gamma_\tau + \Gamma_{had} + N_\nu
\Gamma_\nu
\label{onetwo}
\eeq
where all the data are consistent with the lepton universality expected in the
Standard Model:
$\Gamma_e = \Gamma_\mu = \Gamma_\tau \equiv \Gamma_\ell$, which also predicts
$\Gamma_\nu = 1.991 \pm 0.001 \Gamma_\ell$. We parametrize the total decay
rate of the $Z^0$ into invisible particles as $\Gamma_{inv} =
N_\nu\Gamma_\nu$.
The partial decay rate for leptons is often parametrized as $R_\ell \equiv
\Gamma_{had}/\Gamma_\ell$, and the $Z^0$ decays into heavier quarks by $R_{b,c}
\equiv \Gamma_{b,c/had}$.

Other precision measurements on the $Z^0$ peak~\cite{Zpeak} include
forward-backward asymmetries $A_{FB}$: at the tree level for $f \not= e$:
\beq
{d\sigma\over d \cos \theta}~(e^+e^- \rightarrow \bar ff) \simeq
(1+\cos^2\theta ) ~ F_1 + 2\cos\theta F_2
\label{onethree}
\eeq
and
\beq
A_{FB} = {\left(\int^1_0 - \int^0_{-1}\right) d\cos\theta \cdot {d\sigma\over
d(\cos\theta)} \over
\int^1_{-1} d\cos\theta\cdot{d\sigma\over d(\cos\theta)}} = {3F_2\over 4F_1}
\label{onefour}
\eeq
which takes the value $A_{FB} = {3\over 4} (1-4\sin^2\theta_W)^2$ for
$\mu^+\mu^-$ and $\tau^+\tau^-$ in the Standard Model. Also of interest is the
final-state $\tau$ polarization
\beq
P_\tau = {2(1-4\sin^2\theta_W)\over 1+(1-4\sin^2\theta_W)}
\label{onefive}
\eeq
at the tree level in the Standard Model, whose different functional dependence on
$\sin^2\theta_W$ gives it greater sensitivity. Of particular interest at the SLC
is the difference between the cross-sections for left- and right-handed electron
beams, the polarized-beam asymmetry
\beq
A_{LR} = {\sigma_L - \sigma_R\over \sigma_L + \sigma_R} =
{2(1-4\sin^2\theta_W)\over 1+(1-4\sin^2\theta_W)^2}
\label{onesix}
\eeq
at the tree level. Clearly, this requires a longitudinally-polarized beam, which
SLAC has available,  whereas LEP only has transversely-polarized beams, whose
utility will become apparent shortly.

Table 1 includes a summary~\cite{LEPEWWG} of the current status of all these
key precision
electroweak measurements, among others including the measurements of
$m_W$, which are denominated by those at FNAL~\cite{FNALMW}. We note in
particular the effective number of light neutrino species
\beq
N_\nu = 2.992 \pm 0.011
\label{oneseven}
\eeq
which is an important piece of information for cosmological nucleosynthesis,
discussed here by Keith Olive~\cite{Olive}. Some of the fun physics
involved in the
precise determination of $M_Z$ and $\Gamma_Z$ merits further discussion.

It has been known for some years that the LEP beam energy is sensitive to
terrestrial tides~\cite{tides}. These cause the rock in which LEP is
embedded to rise and
fall by about 25 cm each day, inducing variations in the LEP circumference
$\Delta C \simeq$ 1 mm. Because the angular rotational velocity of the beams is
fixed by the frequency of the RF system, the beams are forced to change
trajectory so as to ``cut corners" when LEP expands, which induces a fractional
change in energy
$\Delta E/E \simeq 10^{-4}$. This tidal effect can be predicted quite reliably,
and measurements of the LEP beam energy, using resonant depolarization of the
transversely-polarized beams, agree well with the predictions of the tidal
model, as seen in Fig. 3a. Since these variations in the beam energy are of order
10 MeV, they need to be taken into account in order to extract $m_Z$, whose
quoted error of about 2 MeV~\cite{LEPEWWG} is much smaller.

\begin{table}
\begin{center}
\caption{}
\vspace*{0.3cm}
\begin{tabular}{|l|rcl|}  \hline
 & \multicolumn{3}{c|}{Measurement}  \\ \hline
$m_Z$ [GeV] & 91.1863 & $\pm$ & 0.0019 \\
$\Gamma_Z$ [GeV] & 2.4947 & $\pm$ & 0.0026 \\
$\sigma^0_{hadr}$ [nb] & 41.489 & $\pm$ & 0.055 \\
$R_1$ &20.783 & $\pm$ & 0.029 \\
$A_{fb}^{0,1}$ & 0.0177 & $\pm$ & 0.0010 \\
$A_\tau$ & 0.1401 & $\pm$ & 0.0067 \\
$A_e$ & 0.1382 & $\pm$ & 0.0076 \\
$\sin^2\theta^{lept}_{eff}$ & 0.2322 & $\pm$ & 0.0010 \\
$R_b$ & 0.2177 & $\pm$ & 0.0011 \\
$R_c$ & 0.1722 & $\pm$ & 0.0053 \\
$A^{0,b}_{fb}$ & 0.0985 & $\pm$ & 0.0022 \\
$A^{0,c}_{fb}$ & 0.0735 & $\pm$ & 0.0048 \\
$A_b$ & 0.897 & $\pm$ & 0.0047 \\
$A_c$ & 0.623 & $\pm$ & 0.085 \\
$\sin^2\theta^{lept}_{eff}$ & 0.23055 & $\pm$ & 0.00041 \\
$1 - m^2_W/m^2_Z$ & 0.2244 & $\pm$ & 0.0042 \\
$m_W$ [GeV] & 80.37 & $\pm$ & 0.08\\
$m_t$ [GeV] & 175.6 & $\pm$ & 5.5 \\
\hline
\end{tabular}
\end{center}
\end{table}

To arrive at this precision, very careful monitoring of the LEP beam energy is
necessary, and several other bizarre effects have shown up. Figure 3b shows
variations in the length of one arc of LEP, which is correlated with the beam
energy by the ``corner-cutting" effect mentioned above. As can be seen in Fig.
3b, a large part of the variation in 1993 is correlated with the height of the
water table in the Jura mountains. This is understandable, since
the rock expands
as it absorbs more water. However, this is not the whole story.
There is also a correlation with changes in the water level of Lake
Geneva~\cite{Lake}, as
seen in Fig. 3c. This reflects the fact that water is run off during the first
part of the year to make room in the Spring for molten snow from the mountains.
With the weight of the water removed, the bedrock rises on a time scale of about
100 days
\footnote{For comparison, we recall that  the North of
Canada is still rising after the last Ice Age.}, causing LEP to expand and its
energy to vary.

\begin{figure}
\hglue.3cm
\mbox{\epsfig{figure=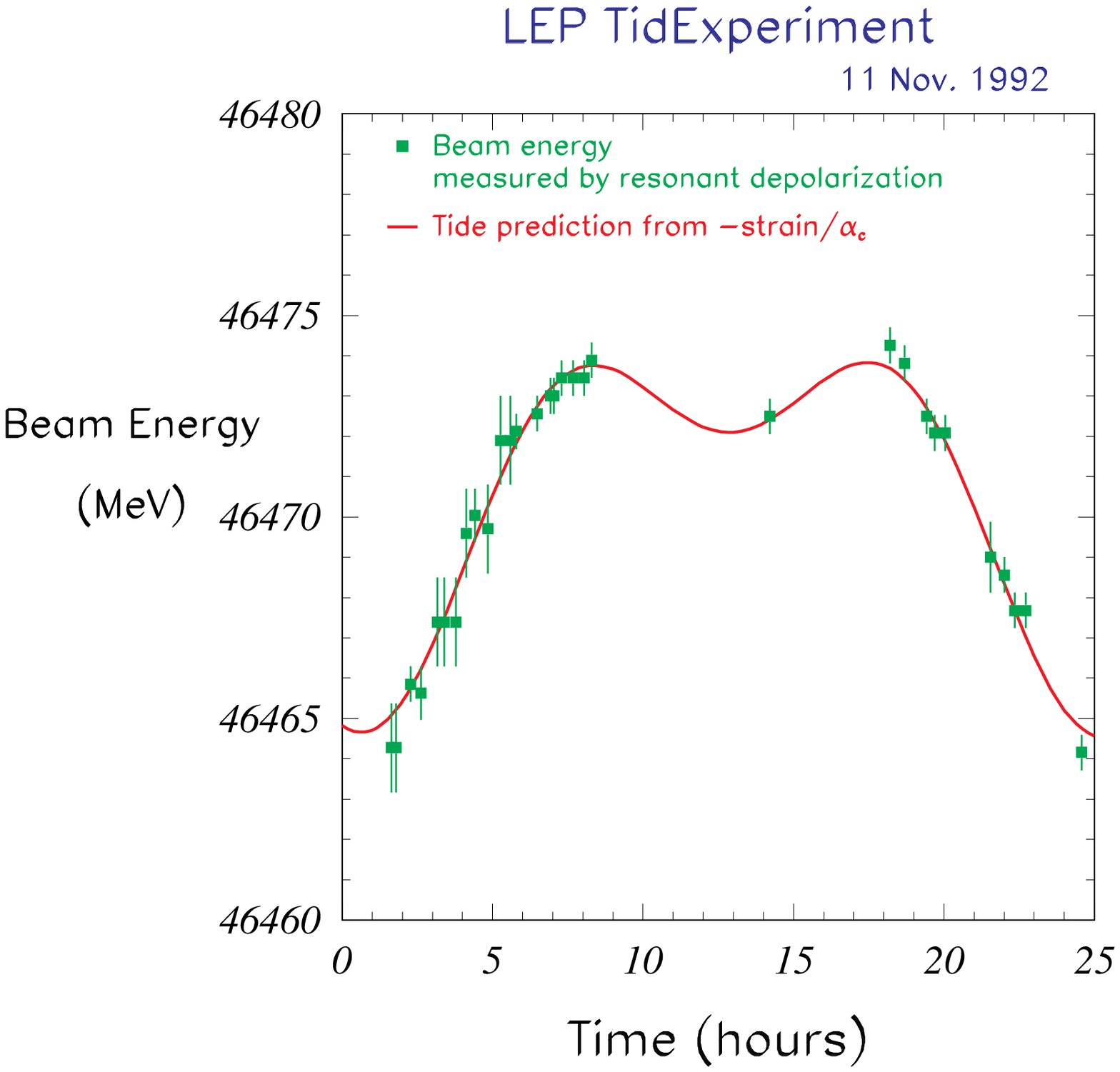,width=5cm}(a)
\epsfig{figure=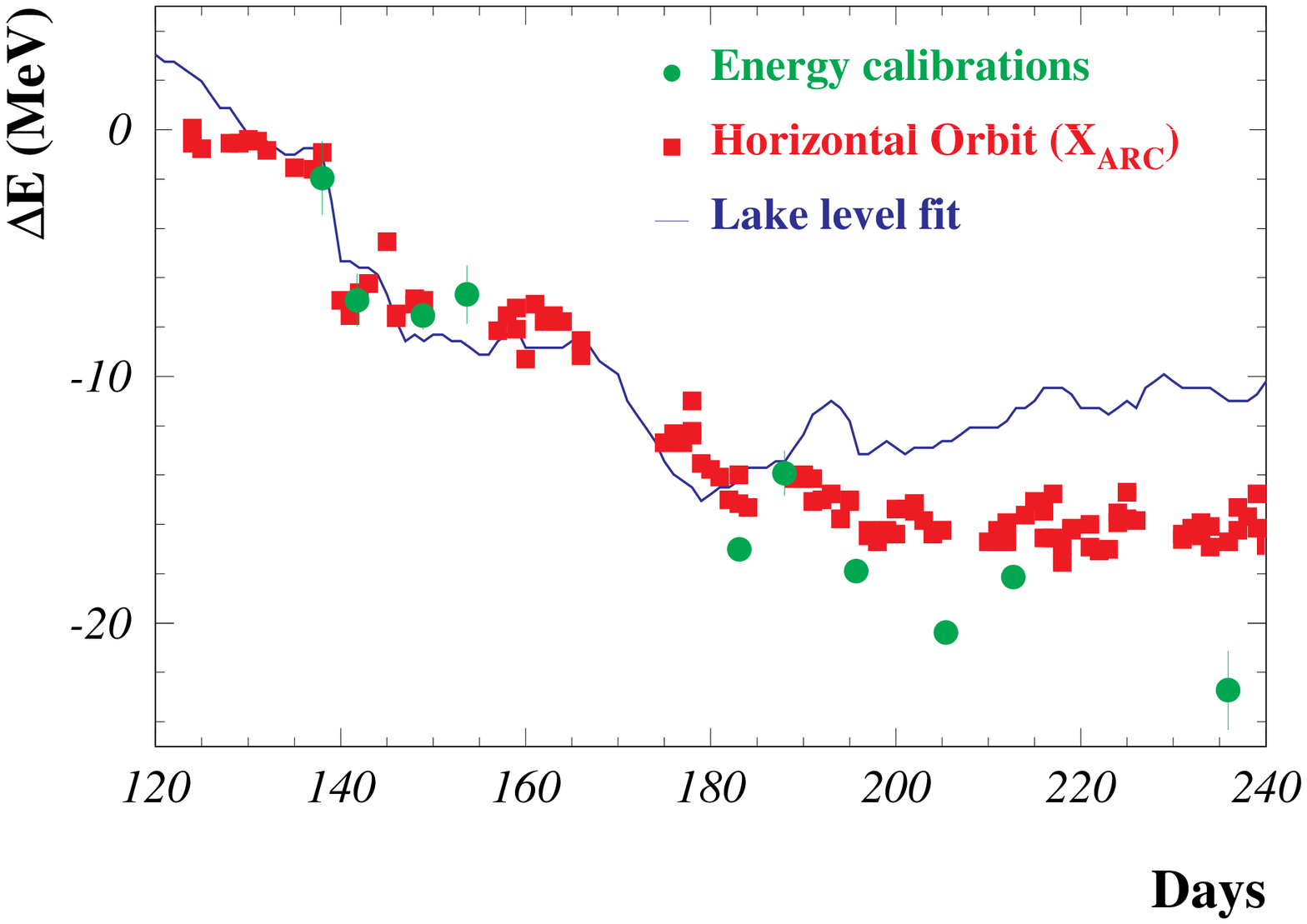,width=5cm}(c)}
\hglue.3cm
\mbox{\epsfig{figure=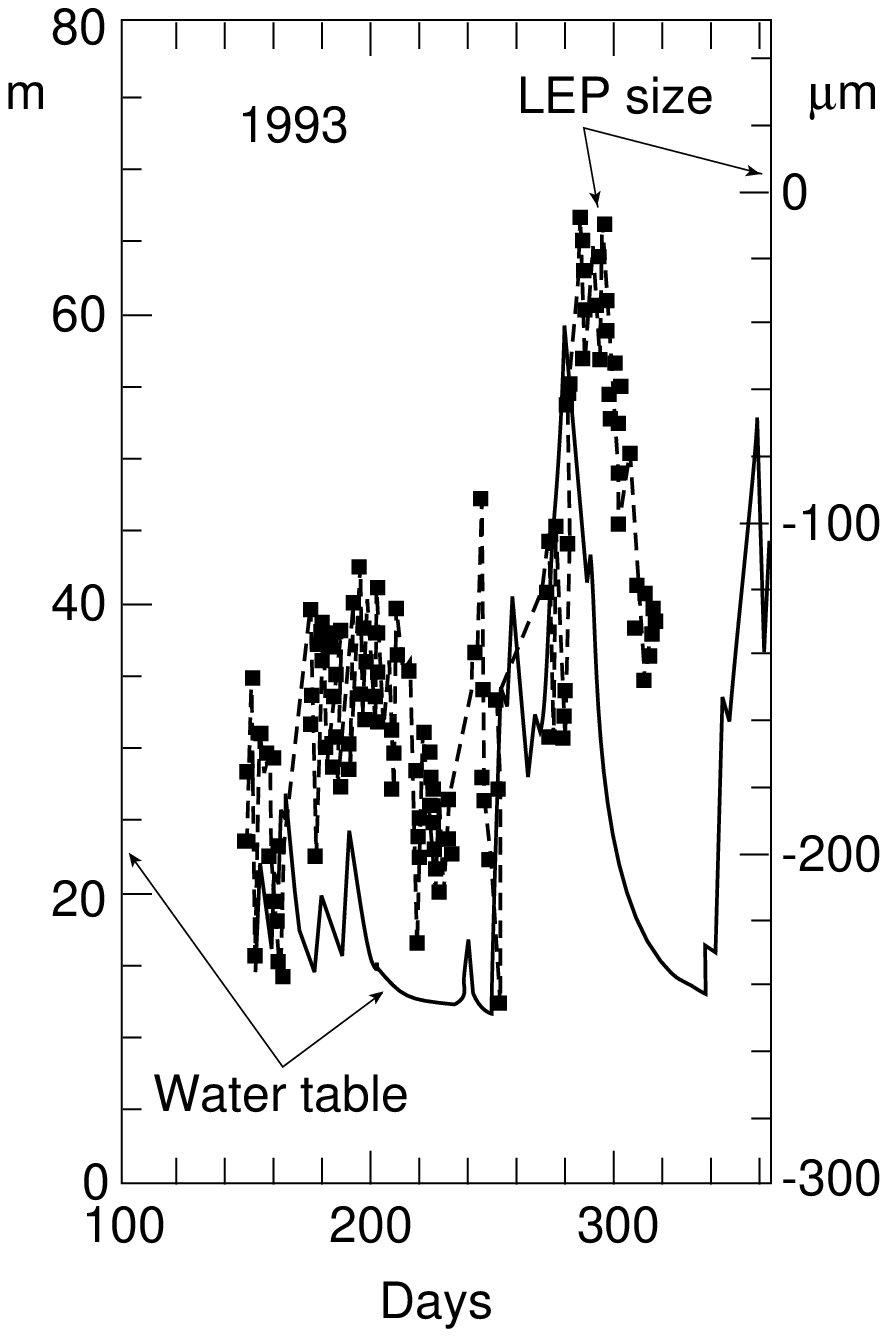,width=5cm}(b)
\epsfig{figure=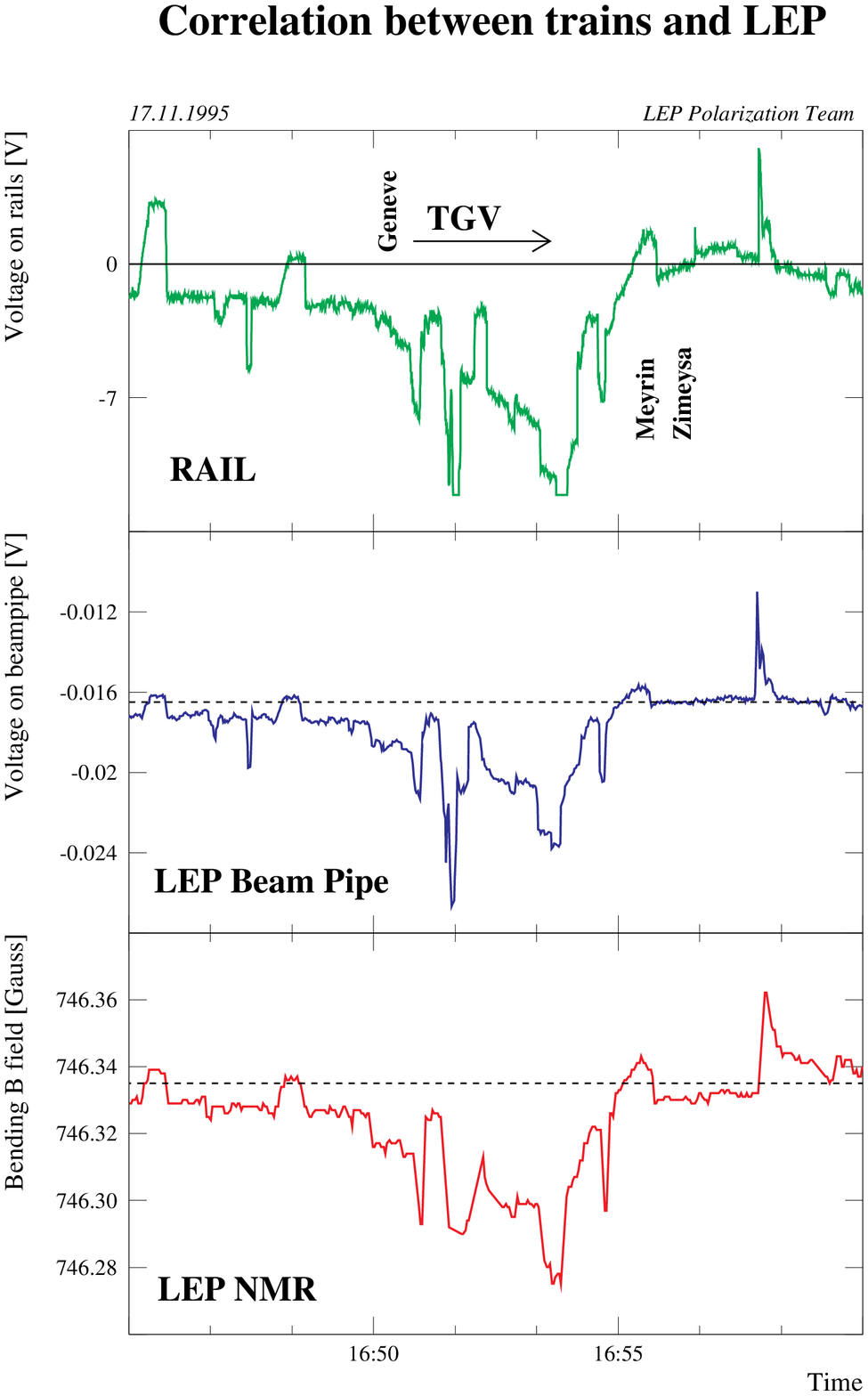,width=5cm}(d)}
\caption[]{(a) Correlation of the LEP beam energy with tides. (b)~Correlation of
the LEP ring size with the height of the water table in the Jura mountains.
(c)~Correlation of the LEP beam energy with the orbit size and the water level
in Lake
Geneva. (d)~Correlation between LEP beams and the passage of a TGV train.}
\end{figure}

In principle, beam energy calibration using resonant depolarization during the
LEP fill monitors all these effects so that they no longer affect the
determination of $m_Z$. However,  more recently it has been
discovered that the beam energy varies systematically during a fill, which is
problematic in view of the fact that the beam energy is typically calibrated at
the end of a fill. This variation is least important
during the night. After some puzzlement, during which possible sources of
electrical interference were sought on the CERN site, the origin was finally
identified as trains on the nearby railway line between Geneva and
France~\cite{TGV}, as
seen in Fig.~3d. The interpretation is that some current leaks from the rails
through the earth (which is not a perfect insulator) and particularly the LEP
ring (which is a much better conductor), before returning to the rails via the
Versoix river. This varying current perturbs the LEP magnets,
which settle into a domain configuration with slightly higher field, and hence
beam energy, as can be seen in Fig. 3d. This effect is potentially larger than the
quoted error on $m_Z$ and has had to be taken into account. Figure 4 tabulates the
latest determinations of
$m_Z$ by the four LEP collaborations~\cite{LEPEWWG}, including this ``TGV
effect".

\begin{figure}
\hglue2cm
\epsfig{figure=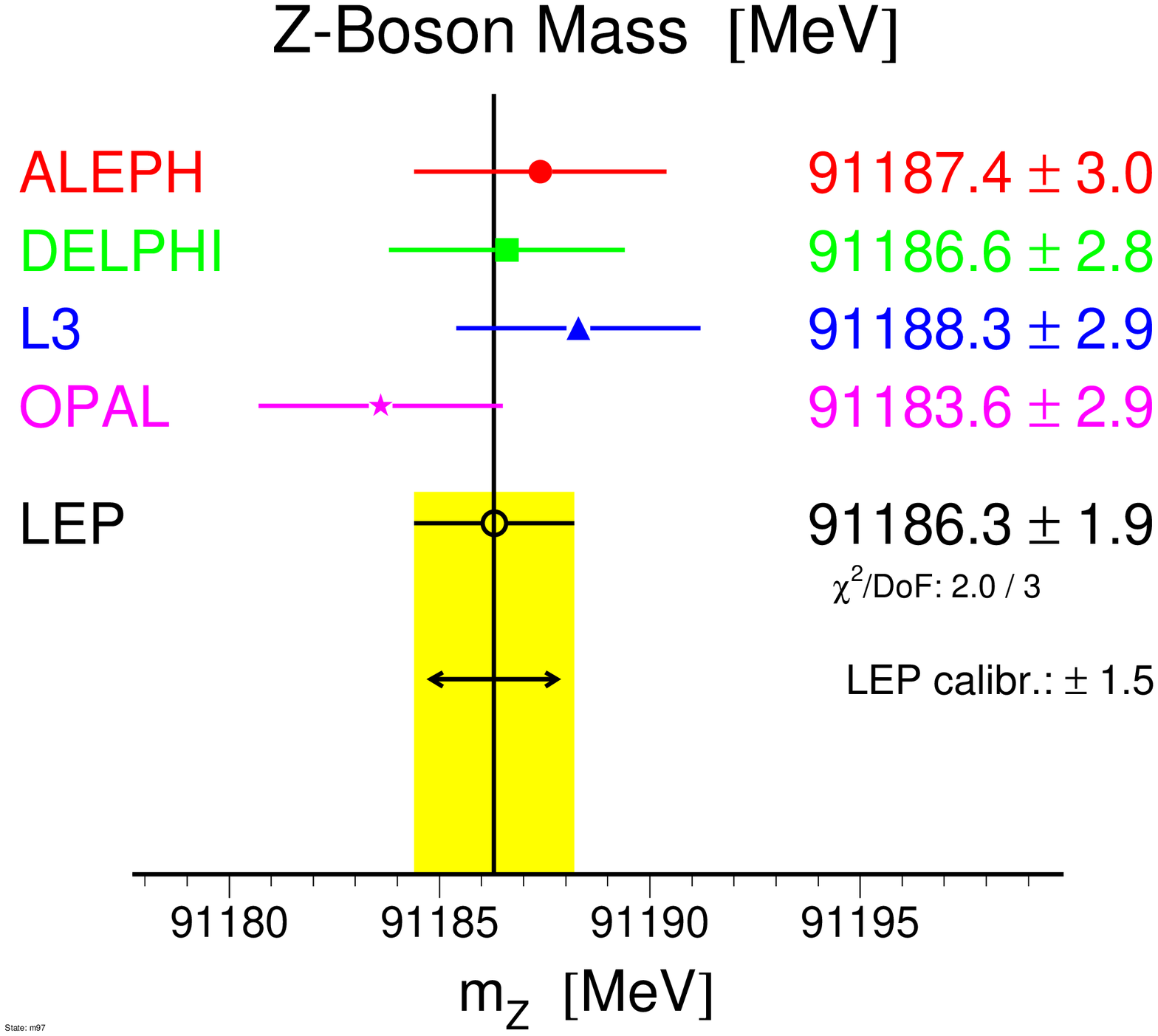,width=8cm}
\caption[]{
LEP measurements of $m_Z$, including the TGV effect.}
\end{figure}

What is the significance of this measurement? At the tree level, in the Standard
Model one has
\beq
m_Z = \sqrt{{\pi\alpha\over\sqrt{2} G_\mu}}~~{1\over\sin\theta_W\cos\theta_W}
\label{oneeight}
\eeq
where $\sin^2\theta_W$ is related to the ratio of gauge boson couplings:
\beq
\sin^2\theta_W = {g^{\prime 2}\over g^{\prime 2} + g^2_2}
\label{onenine}
\eeq
whose value is a key discriminant between different grand unified theories. The
error in $m_Z$ is now comparable to that in the $\mu$ decay constant $G_\mu$.
The value of the fine-structure constant $\alpha$ is well known in the
low-energy Thomson limit, but the value most relevant to $m_Z$ is that at high
energies, which is modified by radiative corrections. These are sensitive to
virtual particles such as the top quark and our quarry in this lecture, the Higgs
boson, whose effects we discuss in the next section.

\subsection{Indirect Indications on $m_H$}
At the one-loop level, using the mass-shell definition of $\sin^2\theta_W$, Eq.
(\ref{onenine}) is modified to become
\beq
m^2_W\sin^2\theta_W = m^2_Z\cos^2\theta_W\sin^2\theta_W =
{\pi\alpha\over\sqrt{2} G_\mu}~~(1+\Delta r)
\label{oneten}
\eeq
The one-loop quantum correction receives important contributions from the top
quark~\cite{Veltmant}:
\beq
\Delta r \ni {3G_\mu\over 8\pi^2\sqrt{2}}~~m^2_t + \ldots
\label{oneeleven}
\eeq
for $m_t \gg m_b$. The divergence in (\ref{oneeleven}) for large $m_t$ reflects
the fact that, without the top quark to complete the doublet started by the
bottom quark, the gauge invariance of the Standard Model would be lost, and with
it renormalizability and predictivity. Likewise, the Higgs boson plays an
essential r\^ole in the spontaneous symmetry-breaking mechanism for generating
particle masses, and the Standard Model would also be non-renormalizable in its
absence. Thus we also expect a Higgs contribution  to $\Delta r$ that is
divergent for large $m_H$. At the one-loop level, this is only
logarithmic~\cite{VeltmanH}: \beq
\Delta r \ni {\sqrt{2}G_\mu\over 16\pi^2}~m^2_W~~\left\{ {11\over 3} \ln
{m^2_H\over m^2_W} + \ldots\right\}
\label{onetwelve}
\eeq
for $m_H \gg m_W$, though numerically less important quadratically-divergent
terms appear at higher-loop level.

Among other quantum-correction effects, we note that the ratio $\rho =
1+\Delta\rho + \ldots$ of low-energy neutral- and charged-current events in
deep-inelastic $\nu$ scattering also depends quadratically on $m_t$, as does
$Z^0\rightarrow \bar bb$ decay~\cite{Zpeak}:
\beq
{\Delta\Gamma_b\over\Gamma^0_b} \ni {-20\over 13}~~{\alpha\over\pi}~~\left[
{m^2_t\over m^2_W} + \ldots \right]
\label{onethirteen}
\eeq
Moreover, much is known about the leading-order radiative corrections beyond one
loop, for example the correction in (\ref{oneten}): $1+\Delta r\rightarrow
1/((1-\Delta r)$, and~\cite{Vanderbij}
\beq
\Delta\rho = {3 G_\mu\over 8\pi^2\sqrt{2}}~~m^2_t~~\left[1-(2\pi^2 - 19)~~{G_\mu
m^2_t\over 8\pi^2\sqrt{2}} + \ldots\right]
\label{onefourteen}
\eeq
These are also included in the codes used to analyze precision
electroweak data~\cite{Dima}.

Such loop corrections also appear in all other electroweak observables. For
example, Fig. 5 compiles the latest determinations~\cite{LEPEWWG} of
$\Gamma_Z$ and shows how
they compare with the Standard Model prediction as a function of
$m_t$, $m_H$ and
the strong gauge coupling $\alpha_s$. We see that the data have the potential to
predict $m_t$, with some uncertainty due to $m_H$ and $\alpha_s$, that may be
reduced by combining many different precision electroweak measurements. This
feature is visible in Fig. 6, which compiles different determinations of the
effective value of $\sin^2\theta_W$ on the $Z^0$ peak [which is closer to the
$\overline{\rm MS}$ definition than to the mass-shell definition used in
(\ref{oneten})].

\begin{figure}
\hglue2cm
\epsfig{figure=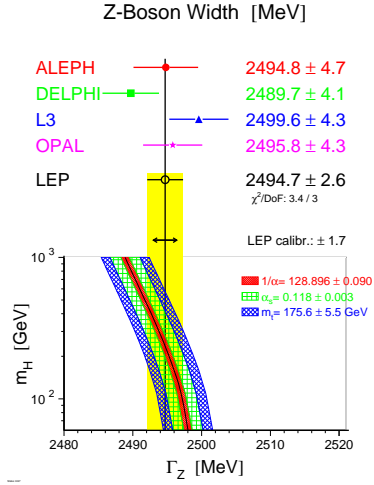,width=5cm}
\caption[]{
Measurements of the $Z^0$ width, including allowance for the TGV effect, compared
with the Standard Model prediction as a function of $\alpha , \alpha_s, m_t$ and
$m_H$.}
\end{figure}

It was pointed out before LEP started operation that $m_t$ could be predicted on
the basis of precision electroweak data as soon as $m_Z$ got measured
precisely~\cite{EF},
and predicting $m_t$ has since become a major industry~\cite{LEPEWWG}.
According to our latest analysis~\cite{EFL96}, the best estimate of $m_t$,
based on the data available since the summer of 1996 and treating $m_H$ as a
free parameter, is

\begin{figure}
\hglue2cm
\epsfig{figure=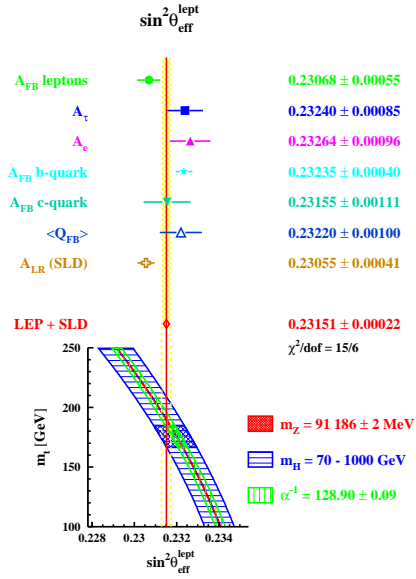,width=8cm}
\caption[]{
Different measurements of the effective $\sin^2\theta_W$ at the $Z$ peak.}
\end{figure}

\beq
m_t = 157^{+16}_{-12}~~{\rm GeV}
\label{onefifteen}
\eeq
This is to be compared with the latest measurements by CDF and $D0$ at
Fermilab, which yield $m_t = 175\pm 6$ GeV~\cite{FNALmt}. The good agreement
between the
indirect estimate and the direct measurement entitles one to combine them, as
seen in Fig. 7, yielding $m_t = 172 \pm 6$ GeV.

Even in the absence of the direct measurement of $m_t$, the precision
electroweak data alone provide~some information about $m_H$, preferring a
central value around 100 GeV~\cite{EFL96,LEPEWWG}. This estimate is
sharpened if the
direct measurement of $m_t$ is included in the fit, yielding the
estimate~\cite{EFL96}
\beq
m_H = 145^{+164}_{-77}~{\rm GeV}~~\left[ \log ({m_H\over{\rm GeV}})
= 2.16 \pm 0.44\right]
\label{onesixteen}
\eeq
within the framework of the Standard Model.

\begin{figure}
\hglue3cm
\epsfig{figure=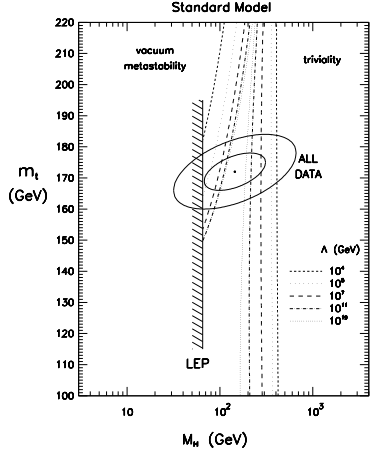,width=5cm}
\caption[]{
Global fit to the precision electroweak data and Fermilab measurements of $m_t$,
compared with the LEP lower limit on $m_H$ and the range expected if the Standard
Model remains unmodified up to a scale $\Lambda$.}
\end{figure}

The range (\ref{onesixteen}) can be compared with the upper limit $m_H \lappeq$
1 TeV suggested by tree-level unitarity~\cite{MHupper}, and the ranges
allowed if the Standard
Model couplings are required to remain finite if run up to some scale $\Lambda$
using the renormalization group equations~\cite{MHrg}. For $\Lambda
\sim_{GUT}$ or $m_P$,
one requires $m_H \lappeq$ 200 GeV, which is relaxed to $m_H \lappeq$ 650 GeV if
one takes $\Lambda = m_H$ itself, as in a lattice
calculation~\cite{MHlattice}.
There are also lower bounds on $m_H$ which follow from examining the behaviour
of the Standard Model effective Higgs potential, and requiring that the standard
electroweak vacuum be stable against transitions to any other state with $\vert
H\vert \leq \Lambda$, or at least metastable with a lifetime longer than the age
of the Universe~\cite{Sher}.

These bounds are compared in Fig. 7 with the range (\ref{onesixteen}) favoured
by the experimental measurements. We see that there is no indication of any
breakdown of the Standard Model at any scale $\Lambda \lappeq m_P$. However, we
nevertheless draw some encouragement for possible physics beyond the Standard
Model, as discussed in the next section.

\subsection{Motivations for Supersymmetry}

The primary theoretical motivation for the appearance of
supersymmetry~\cite{susy} at
accessible energies is to understand the origin of the large hierarchy of mass
scales in physics~\cite{hierarchy}: how and why is $m_W$ so much smaller
than $m_P$, the only
candidate we have for a fundamental mass scale in physics? This question is made
particularly acute by radiative corrections, which make such a hierarchy seem
very unnatural. It is one thing to derive or postulate the existence of a very
small bare mass parameter. It is another if radiative corrections to this bare
quantity are very large, so that a small physical value can only be obtained at
the price of an apparently conspiratorial cancellation between (almost) equal
and opposite  large bare and quantum contributions: the ``fine-tuning" problem.
This problem is not too acute for fermion masses $m_f$, whose one-loop
correction has the form
\beq
\delta m_f = 0\left({\alpha\over\pi}\right)m_f\ln \left({\Lambda\over
m_f}\right)
\label{oneseventeen}
\eeq
where $\Lambda$ is some cut-off representing an energy scale where the physics
gets modified, which might be $m_P$ at most. Because the divergence in
(\ref{oneseventeen}) is only logarithmic, this correction is not much larger
than the physical value of $m_f$, reflecting the fact that its smallness is
safeguarded by chiral symmetry and poses no serious problem of naturalness or
fine tuning.

This is, however, a problem for an elementary scalar boson, such as the Higgs
boson of the Standard Model, whose mass must be within an order of magnitude of
$m_W: m_W = 0\left(\sqrt{{\alpha\over\pi}}\right)^{0\pm 1} m_H$. Each of the
diagrams in Fig. 8a contributes a quadratically-divergent radiative correction
\beq
\delta m^2_H \simeq g^2_{f,W,H}~~\int^{\Lambda} {d^4k\over
(2\pi)^4}~~{1\over k^2} \simeq 0\left({\alpha\over\pi}\right) \Lambda^2
\label{oneeighteen}
\eeq
which is many orders of magnitude larger than the physical value of $m_H$ if we
take $\Lambda \sim m_P$ or $m_{GUT}$~\cite{hierarchy}.

\begin{figure}
\hglue3cm
\epsfig{figure=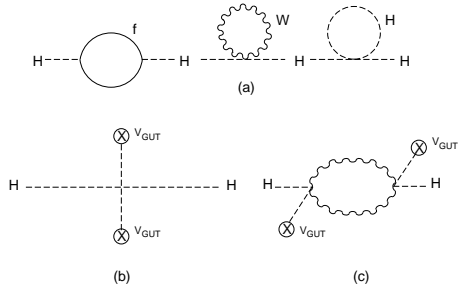,width=6cm}
\caption[]{
Potentially large contributions to $m_H$ from (a) quadratic divergence in the
Standard Model, (b) couplings to GUT Higgs bosons, and (c) logarithmic corrections
to the latter.}
\end{figure}

The remedy~\cite{hierarchy} offered by supersymmetry~\cite{susy} is based on
the fact that boson and fermion
loops have opposite signs. Therefore, if one has equal numbers of bosons and
fermions, and if their couplings are equal, the quadratically-divergent
corrections will cancel:
\beq
\delta m^2_{W,H} \simeq -\left({g^2_F\over 4\pi}\right)~(\Lambda^2 + m^2_F)
+ \left({g^2_B\over 4\pi}\right)~(\Lambda^2 + m^2_B) ~\simeq~
0\left({\alpha\over\pi}\right)~~( m^2_B - m^2_F)
\label{onenineteen}
\eeq
with a possible logarithmic factor, which is acceptably small $(\lappeq
m^2_{W,H})$ if
\beq
\vert m^2_B - m^2_F\vert \lappeq 1~~{\rm GeV}^2
\label{onetwenty}
\eeq
Approximate supersymmetry also removes the threat of the GUT radiative corrections
in Fig. 8c, though it does not by itself explain why the couplings in Fig. 8b should
vanish.
This naturalness argument provides the only theoretical argument why
supersymmetry should appear at low energies, rather than (say) at $m_P$ where it
is apparently required for the consistency of string theory.

It should be emphasized, though, that this naturalness argument is qualitative,
and rather a matter of taste. How much fine tuning of bare and one-loop masses
is one prepared to tolerate: a factor of 2? 10? 100?  Moreover, the Standard
Model is mathematically consistent, in the sense that it is renormalizable and
hence calculable, however much fine tuning there may be. The fine-tuning
argument~\cite{hierarchy} is essentially one of physical intuition.

In the minimal supersymmetric extension of the Standard Model
(MSSM)~\cite{MSSM}, all the
known particles are promoted to doublets ($L,  Q$ for lepton and quark doublets,
$E^c, U^c, D^c$ for charged lepton and quark singlets) with identical internal
quantum numbers, but spins differing by half a unit. The supersymmetric part of
the Lagrangian is determined by the gauge  interactions, which are identical
with those of the Standard Model, and by the superpotential:
\beq
W = \sum_{L,E^c}~ \lambda_L ~ LE^c H_1 + \sum_{Q,U^c}~\lambda_U ~ QU^c H_2 +
\sum_{Q,D^c} ~ \lambda_D ~ QD^c H_1 + \mu H_1H_2
\label{onetwentyone}
\eeq
The first three terms yield Yukawa couplings that give masses to the charged
leptons, charge 2/3 and charge -1/3 quarks, respectively. Two Higgs doublets
$H_{1,2}$ are needed to provide all the masses, in order to cancel triangle
anomalies and for the superpotential to be holomorphic. In addition to the
Yukawa interactions, the superpotential (\ref{onetwentyone}) provides, together
with the gauge interactions, quartic self interactions of the scalar components
of the supermultiplets, enabling predictions to be made for the masses of our
quarries in this lecture, the physical Higgs bosons in the MSSM, as we shall
discuss later.

In addition to the supersymmetric parts of the Lagrangian of the MSSM, there
must be terms that break supersymmetry, providing in particular masses for the
(as yet) unseen supersymmetric partners of the particles of the MSSM, such as
the scalars: $m_{0_i}$ and gauginos $M_\alpha$. If these parameters are much
larger than $m_W$, the latter becomes very sensitive to the choices of input
parameters, and the fine-tuning problem returns. How large the
$(m_{0_i},M_\alpha)$ may be depends on the degree of fine tuning $\eta$ that one
is comfortable with~\cite{EENZ,BG}:
\beq
{\Delta m_W\over m_W} \leq \eta~{\Delta ({\rm input})\over({\rm input})}
\label{onetwentytwo}
\eeq
Figure 9 shows estimates of some sparticle masses obtained by requiring
$\eta~\leq~10$. We see that such an analysis favours
relatively light sparticle
masses: $m_{0_i}, M_\alpha \lappeq$ few hundred GeV, whether or not the scalar
mass parameters $m_{0_i}$ are assumed to be the same for all three generations of
squarks~\cite{DG}. However, this is not a rigorous upper bound: in some
sense, a more
characteristic prediction of the MSSM is provided by the mass of the Higgs boson,
as we discuss shortly.

\begin{figure}
\hglue2cm
\epsfig{figure=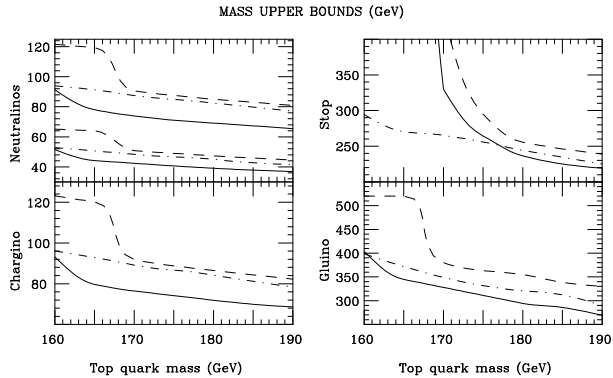,width=8cm}
\caption[]{
Upper bounds on sparticle masses, applying the condition (1.22) with $\eta = 10$,
for universal scalar masses (dot-dashed lines) and non-universal masses (dashed
lines). The solid lines are in the former case, neglecting loop corrections to
Higgs boson masses.}
\end{figure}

To my mind, the precision electroweak data currently provide two tentative
(s)experimental motivations for low-energy supersymmetry~\cite{susyGUT}.
One is the consistency
of measurements of the gauge coupling strengths of the Standard Model with the
hypothesis of supersymmetric grand unification. As seen in Fig. 10, when viewed
on a scale from 0 to 1, the predictions of both supersymmetric and
non-supersymmetric GUTs compare
very well with the precision electroweak measurements discussed earlier. Blowing
the scale up by a factor 10 reveals a significant discrepancy with the
non-supersymmetric GUT prediction, whereas the supersymmetric GUT prediction is
satisfactory. Blowing the scale up by a further factor of 10 indicates that the
sparticle masses cannot be exactly $m_Z$, but, to my mind, the electroweak data do
not permit an interestingly precise indirect determination of the supersymmetric
threshold. The second tentative experimental indication in favour of low-energy
supersymmetry is provided by the apparent preference~\cite{EFL96,LEPEWWG}
for a light Higgs boson
manifested by the precision electroweak data which was discussed earlier, and is
consistent with the prediction of the MSSM.

\begin{figure}
\hglue2cm
\epsfig{figure=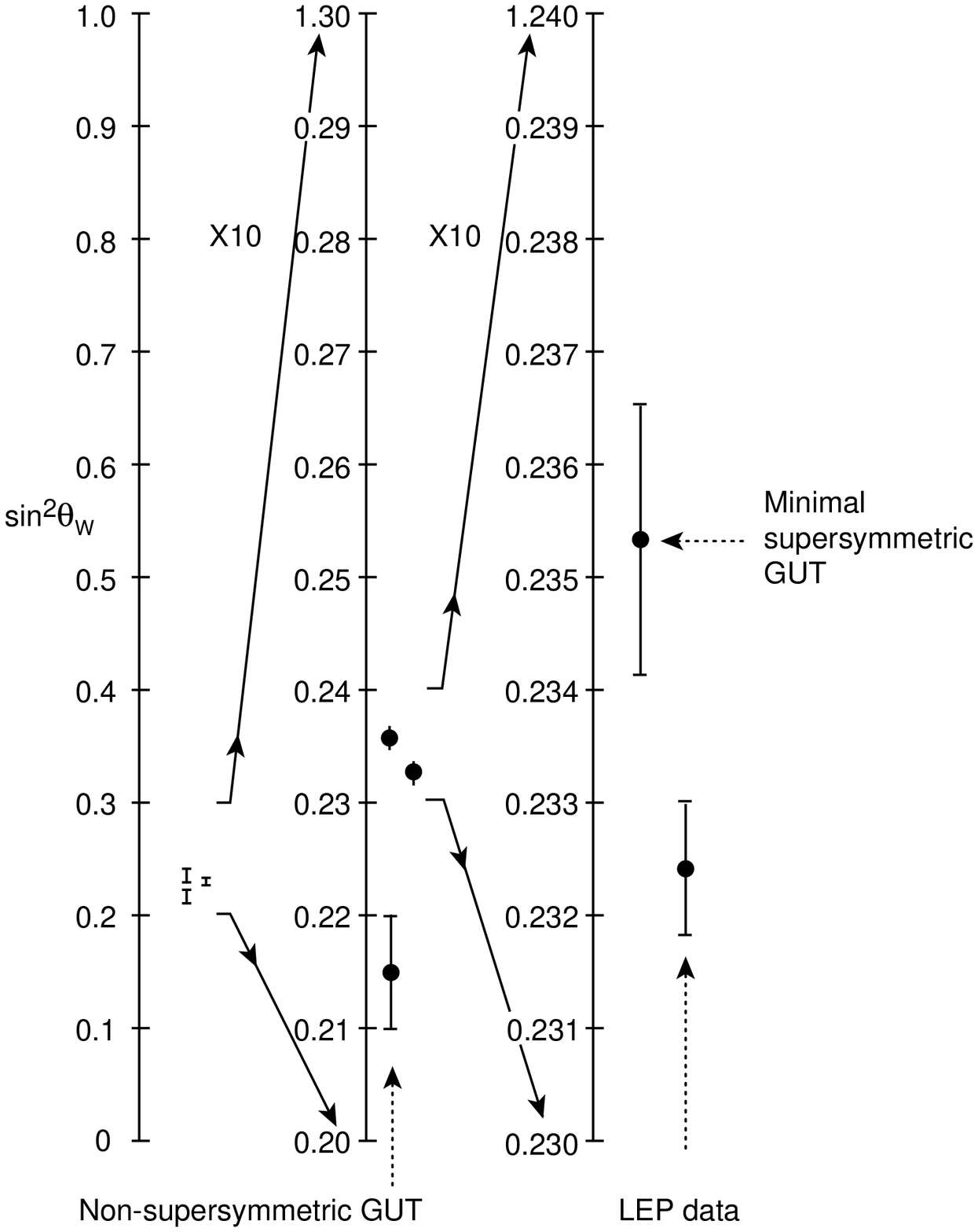,width=8cm}
\caption[]{
Measurements of $\sin^2\theta_W$: note that the non-supersymmetric GUT prediction
disagrees significantly with the data. The minimal supersymmetric GUT prediction
assumes unrealistically that all sparticles have masses $m_Z$. Realistic spectra
give predictions in agreement with the data.}
\end{figure}

Since the MSSM contains two complex Higgs doublets, it contains eight real Higgs
degrees of freedom, three of which become the longitudinal polarization states
of the $W^\pm$ and $Z^0$, leaving five physical states. Three of these are
neutral - two scalars $h, H$ and one pseudoscalar $A$ - and two are charged Higgs
bosons $H^\pm$. At the tree level, all the MSSM Higgs masses and couplings are
determined in terms of two parameters, which may be taken as the pseudoscalar
mass $m_A$ and the ratio of Higgs v.e.v.'s: $\tan\beta\equiv {v_2 / v_1}$.
For example, one has~\cite{MSSM}
\beq
m^2_{h,H} = {1\over 2}~\left[m^2_A + m^2_Z\right] \mp
\sqrt{(m^2_A+m^2_Z)^2 - 4m^2_Zm^2_A\cos^2 2\beta}
\label{onetwentythree}
\eeq
from which one sees that $m_h < m_Z$ at the tree level. This lightness reflects
the fact that the quartic Higgs self-coupling in the MSSM is relatively weak,
since it originates from the electroweak $D$ terms and is $0(g^2)$.

However, the upper bound on $m_h$ is significantly relaxed by one-loop radiative
corrections~\cite{delMH}, which take the following form in the limit where
the stop squark masses $m_{\tilde t_{1,2}}$ and $ m_A\gg m_Z$:
\bea
m^2_h &=& m^2_Z \cos^2 2\beta + (\Delta m^2_h)_{ILL} + (\Delta m^2_h)_{mix}
: \nonumber \\
(\Delta m^2_h)_{ILL} &=& {3m^4_t\over 4\pi^2v^2}~\ln \left({m^2_{\tilde t_1}
- m^2_{\tilde t_2} \over m^2_t}\right)~~\left[ 1 + 0\left({m^2_W\over
m^2_t}\right)\right]~,\nonumber \\
(\Delta m^2_h)_{mix} &=& {3m^4_t\over 8\pi^2v^2}~\left[2h (m^2_{\tilde
t_1},m^2_{\tilde t_2}) + \tilde A^2_t f(m^2_{\tilde t_1},m^2_{\tilde
t_2})\right]~~\nonumber \\
&&\left[ 1+0\left(m^2_W\over m^2_t\right)\right]
\label{onetwentyfour}
\eea
where $\tilde A_t \equiv A_t - \mu \cot \beta$ and
\beq
h(a,b) \equiv {1\over a-b},~~f(a,b) \equiv {1\over
(a-b)^2}~~\left[2-\left({a+b\over a-b}\right)\ln \left({a\over
b}\right)\right]
\label{onetwentyfive}
\eeq
One-loop corrections to Higgs coupling vertices are also known, as are the
leading two-loop corrections to the Higgs mass, for which accurate
renormalization-group-improved formulae are known~\cite{improved}.

As seen in Fig. 11, the net effect of the radiative corrections is to increase
the maximal $h$ mass to about 150 GeV. The allowed range for different
$\tan\beta$ is compares well with the range preferred by the precision
electroweak data for different $m_t$ shown in Fig. 7. These correspond to a
probability of 32
\% that the true values of $m_t$ and the Higgs mass lie in the range allowed by
the MSSM, compared to 27 \% for the range allowed by the Standard Model if no new
physics intervenes below the Planck scale~\cite{EFL96}. Supersymmetry is
certainly
consistent with the data, though we cannot yet exclude its absence!

\begin{figure}
\hglue2.5cm
\epsfig{figure=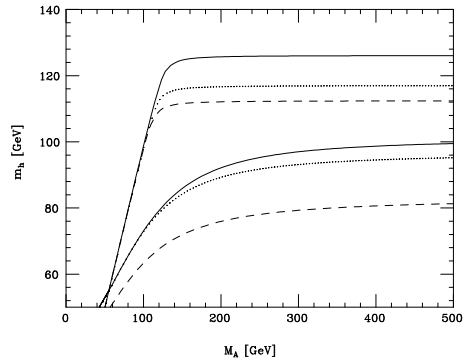,width=6cm}
\caption[]{
The radiatively-corrected mass of the lightest MSSM Higgs boson mass $m_h$ as a
function of $m_A$, for $\tan\beta$ = 1.6 (lower curves) and 15 (upper curves), and
maximal mixing (solid lines), minimal mixing (dashed lines), and intermediate
mixing (dotted lines), assuming $m_t$ = 175 GeV and $m_{\tilde q}$ = 1 TeV.}
\end{figure}

\subsection{Higgs Search at LEP 2}

The dominant mechanism for Higgs production at LEP 2 in the Standard Model is
$e^+e^- \rightarrow Z^0 + H$~\cite{ZH}, whose tree-level cross section is
\beq
\sigma_{ZH} = {G^2_F m^4_Z\over 96 \pi s}~~(v^2_e + a^2_e)\lambda^{1/2}~~
{\lambda + 12 m^2_Z/s \over (1-m^2_Z/s)^2}
\label{onetwentysix}
\eeq
where $a_e = -1$ and $v_e = -1+4\sin^2\theta_W$ are the $Z^0 e^+e^-$ couplings in
the Standard Model, and
$\lambda = (1-m^2_Hs - m^2_Z/s)^2 - 4(m^2_H m^2_Z/s^2)$ is the conventional
two-body phase-space factor. The proper electroweak radiative corrections to
(\ref{onetwentysix}) are small~\cite{LEP2}: $\delta \sigma / \sigma \lappeq$
1.5 \%. More
important are initial-state radiative (ISR) corrections, which in leading order
give
\beq
<\sigma > = \int^1_{x_H} ~dx~G(x)~\sigma(xs)
\label{onetwentyseven}
\eeq
where $x_H \equiv m^2_H/s$ and $G(x)$ is a ``radiator function", which
is known to $0(\alpha^2)$~\cite{LEP2}. In addition, one must allow for
off-shell $Z^0$
production by incorporating finite-width effects. To understand the full reach
of LEP 2 for the Higgs search, one must also take into account the reaction
$e^+e^-\rightarrow\bar\nu_e H \nu_e$, due to $W^+W^-$ fusion~\cite{JP}. All
these effects are included in Fig. 12.

\begin{figure}
\hglue2.5cm
\epsfig{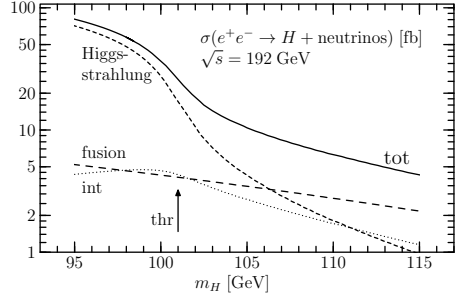}
\caption[]{
Cross section for Higgs boson production at LEP~2, including $e^+e^-\rightarrow
(Z~\rightarrow~\bar\nu\nu)~+~H$ and $e^+e^-\rightarrow \nu_e\bar\nu_e + H$.}
\end{figure}

Since the Higgs couplings to other particles are proportional to their masses,
the dominant Higgs decays
are those into the heaviest available particles (see the first paper
in~\cite{ZH}), notably $H\rightarrow \bar bb$~\cite{LEP2}: \beq
\Gamma (H\rightarrow\bar bb)\simeq {3G_F\over 4\sqrt{2}\pi}~m^2_b~(m_H)~
m_H~\left[ 1 + 5.67 \left({\alpha_s\over\pi}\right) + \ldots \right]
\label{onetwentyeight}
\eeq
to be compared with $H\rightarrow\tau^+\tau^-$:
\beq
\Gamma (H\rightarrow\tau^+\tau^-) \simeq {G_F\over 4\sqrt{2}\pi}~~m^2_\tau~ m_H
\label{onetwentynine}
\eeq
as well as $H\rightarrow \bar cc$ which is given by (\ref{onetwentyeight}) with
$m_b\rightarrow m_c$, and $H\rightarrow gg$ which is dominated by top loops:
\beq
\Gamma (H\rightarrow gg) \simeq {G_F \alpha^2_s~(m_H)\over 36 \sqrt{2}
\pi^3}~m^3_H
\label{onethirty}
\eeq
The decays $H\rightarrow\gamma\gamma$, $W^+W^-$ and $Z^0Z^0$, are not important
at LEP 2, though they are important for LHC Higgs searches~\cite{LHC}. The
total
Higgs decay width $\Gamma_H $ is less than 3 MeV for $m_H <$ 100 GeV.

Each individual LEP experiment has one or two candidate events for Standard
Model Higgs production, but these do not coincide in mass. Individual
experiments limit $m_H \gappeq$ 70 GeV~\cite{MHexp}, and the combined LEP
limit so far is
estimated to be $m_H \gappeq$ 75 to 77 GeV. Eventually, experiments at LEP 2
should be sensitive to $m_H \lappeq E_{cm} - m_Z -$ few GeV, or $m_H \lappeq$
95(100) GeV  if the maximum $E_{cm}$ reaches 192(200) GeV.

In the context of the MSSM, there are two interesting Higgs production
processes: $e^+e^-\rightarrow Z^0h$ and $e^+e^-\rightarrow Ah$.
The former has a tree-level cross section smaller than that for
 $e^+e^-\rightarrow Z^0H$ in the Standard Model by a factor $\sin^2(\beta
-\alpha)$ where $\alpha$ is a mixing angle in the Higgs sector~\cite{LEP2}.
Fortunately,
$\sigma (e^+e^-\rightarrow Ah) \propto \cos^2 (\beta - \alpha )$, so that there
is some complementarity between these two processes, unless $m_A$ is large. The
dominant supersymmetric Higgs decay modes are likely to be similar to those in
the Standard Model, though there is a nightmare possibility that
invisible $hA
\rightarrow \chi\chi$ decays might dominate! Ignoring, for the moment, this
possibility which is in any case disfavoured by the lower limit on $m_\chi$
discussed in Lecture 3, the unsuccessful searches so far for $e^+e^-\rightarrow
Z^0h$ and $Ah$ indicate that $m_h \gappeq$ 63 GeV~\cite{MSSMexp}, whatever
the value of
$\tan\beta$. Because of the radiative correction (\ref{onetwentyfour}),
(\ref{onetwentyfive}), there is no guarantee that LEP 2 will find the lightest
supersymmetric Higgs boson. Fortunately, it seems that the LHC will be able to
complete the coverage of MSSM parameter space.

\section{$W$ Physics}
\setcounter{equation}{0}

\subsection {$W$ Mass}

The world average of direct measurements from  experiments other than those at
LEP yields~\cite{LEPEWWG}
\beq
m_W = 80.37 \pm 0.08~{\rm GeV}
\label{twoone}
\eeq
and is dominated by measurements at the Fermilab $\bar pp$
collider~\cite{FNALMW}. This can be
compared with the theoretical prediction that one makes within the Standard
Model, using a global fit to the precision electroweak data~\cite{LEPEWWG}:
\beq
m_W = 80.323 \pm 0.042~{\rm GeV}
\label{twotwo}
\eeq
A large fraction of this error: (+13, -24) MeV is associated with uncertainty in
the Higgs mass: 60 GeV $< m_H <$ 1 TeV as seen in Fig. 13~\cite{LEP2}. This
arises
principally via the one-loop radiative correction $\Delta r$ to $m_W$:
\beq
G_\mu = {\alpha\pi\over\sqrt{2}}~~ m^2_W ~~(1-m^2_W/m^2_t)~~{1\over 1-\Delta r}
\label{twothree}
\eeq
whose leading $m_H$-dependence was given was Eq. (\ref{onetwelve}).

\begin{figure}
\hglue2cm
\epsfig{figure=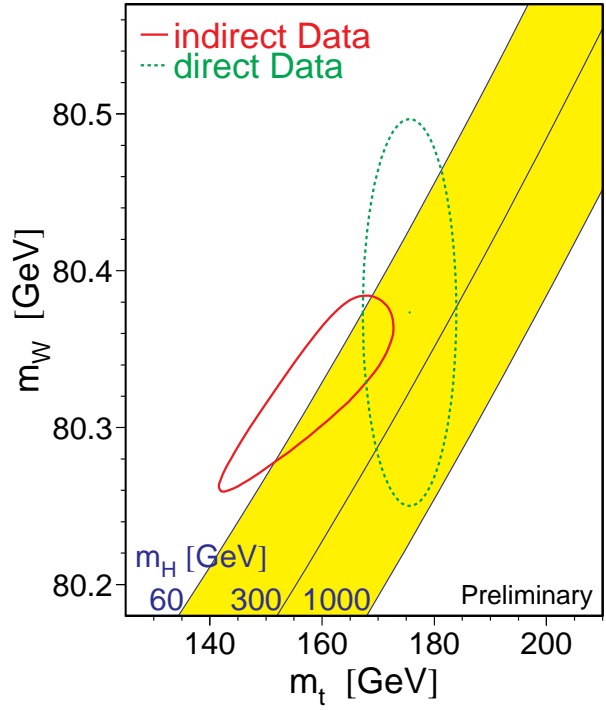,width=8cm}
\caption[]{
Indirect predictions of $m_t$ and $m_W$ (solid line) based on precision electroweak
data from LEP and elsewhere, compared with the present
direct measurements (dashed line) and calculations for different values of $m_H$.}
\end{figure}

A more precise direct measurement of $m_W$ would help constrain $m_H$ within the
Standard Model, as well as constrain possible extensions such as supersymmetry.
 As an example of the possible impact on $m_H$, Table 2
lists the errors in $m_H$ that would result from measurements of $m_W$ with
errors of 25 and 50 MeV, assuming that $m_t$ is measured to be 180 $\pm$
5 MeV~\cite{LEP2}.
We see that a direct measurement of $m_W$ with an error between 25 and 50 MeV,
which seems possible at LEP 2, would have significant impact on the prediction of
$m_H$ within the Standard Model.

\begin{table}
\begin{center}
\caption{Errors in $M_H$ for assumed errors in $m_W$}\vspace*{0.3cm}
\begin{tabular}{|c||c|c|}  \hline
Nominal & \multicolumn{2}{c|}{$\Delta m_W$} \\
\cline{2-3} Value of $m_H$ & 25 MeV & 50 MeV \\ \hline
100 & (+~86, -~54) & (+140, -72) \\
300 & (+196, -126) & (+323, -168) \\ \hline
\end{tabular}
\end{center}
\end{table}

Two methods of measuring $m_W$ at LEP 2 appear to be the most promising. One is
from the threshold cross section for $e^+e^-\rightarrow W^+W^-$, which could
yield~\cite{LEP2}
\beq
\Delta m_W \geq 91~{\rm GeV}~~\sqrt{{100 pb^{-1}\over{\cal L}}}
\label{twofour}
\eeq
where ${\cal L}$ is the accumulated luminosity. The inequality (\ref{twofour})
is saturated at $E_{CM}$ = 161 GeV under the idealized assumptions that $W^+W^-$
events are selected with 100 \% efficiency and no background. Theoretical issues
arising in the calculation of $\sigma (e^+e^-\rightarrow W^+W^-)$ are discussed
in Section 2.2.

Alternatively, one may measure $m_W$ by the direct reconstruction of $W^\pm$
decays, which should yield~\cite{LEP2}
\beq
\Delta m_W \geq {\Gamma_W\over\sqrt{N}} \simeq 50~{\rm MeV}~~\sqrt{{100
pb^{-1}\over {\cal L}}}
\label{twofive}
\eeq
This estimate is valid at any centre-of-mass energy above about 170 GeV, again
under the idealized assumptions of 100 \% efficiency, no background and perfect
detector resolution. In practice, the classes of $W^+W^-$ events that can be
used for such direct reconstruction are $(W^\pm \rightarrow \bar
qq)~(W^\mp\rightarrow \ell\nu)$ and $(W^+\rightarrow \bar qq)~(W^-\rightarrow \bar
qq)$. In the latter case,  questions arise whether the $W^\pm$ hadronize
independently, and whether any interference or collective effects such as colour
reconnection~\cite{colreconn} and Bose-Einstein correlations~\cite{BE}
have a significant impact on the
error (\ref{twofive}) with which $m_W$ may be measured. This problem will be
discussed in Section 2.3 with particular emphasis on our
analysis~\cite{EGZ,EGWW,EGdiff} of the
space-time development of $e^+e^-\rightarrow W^+W^-\rightarrow$ hadrons events.

Finally, it should be noted that one could in principle measure $m_W$ using the
end-point energy of the $W^\pm \rightarrow \ell^\pm \nu$ spectrum:
\beq
\Delta m_W = {\sqrt{S-m^2_W}\over m_W}~~\Delta E_{L^\pm}
\label{twosix}
\eeq
Unfortunately, the end point is so smeared by finite-width and ISR effects, and
so limited in statistics, that it does not appear to be a competitive way to
determine $m_W$~\cite{LEP2}.

\subsection{Cross-Section for $e^+e^-\rightarrow W^+W^-$}

Since the $W^\pm$ are unstable, the $W^+W^-$ final state must be considered in
conjunction with other mechanisms for producing the same four-fermion final
states, which are in principle indistinguishable~\cite{LEP2}. Indeed, any
separation \beq
\sigma_{4f} = \sigma_{W^+W^-} + \sigma_{bkgrd}
\label{twoseven}
\eeq
is not even gauge invariant! The $W^+W^-$ contribution in
(\ref{twoseven}) can be further decomposed in the form:
\beq
\sigma_{W^+W^-} = \sigma^0_{W^+W^-}~(1 + \delta_{ew} + \delta_{QCD}) + \ldots
\label{twoeight}
\eeq
where $\sigma^0_{W^+W^-}$ is the tree-level Born cross-section for producing
off-shell $W^\pm$ via the three ``classic" $\nu -, \gamma -$ and $Z^0$-exchange
``CC03" diagrams shown in Fig.~14,  $\delta_{ew,QCD}$ represent one-loop
electroweak and QCD corrections, respectively and the dots in (\ref{twoeight})
represent higher-order corrections. The off-shell tree-level cross section
$\sigma^0_{W^+W^-}$ may in turn be written as
\beq
\sigma^0_{W^+W^-}(s) = \int^s_0~ds_1
~\int_0^{(\sqrt{s}-\sqrt{s_1})^2}~ds_2~\rho(s_1)~\rho(s_2)~\sigma^0(s,s_1,s_2)
\label{twonine}
\eeq
where
\beq
\rho(s) = {1\over\pi}~{\Gamma_W\over m_W}~{s\over (s-W^2_W)^2 +
s^2\Gamma^2_m/m^2_W}
\label{twoten}
\eeq
is a relativistic Breit-Wigner spectral function with a mass-dependent width
$\Gamma_W(s) = s\Gamma_W/m^2_W$.

\begin{figure}
\hglue3.5cm
\epsfig{figure=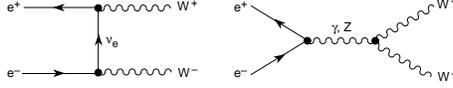,width=6cm}
\caption[]{
Lowest-order CC03 diagrams for $e^+e^-\rightarrow W^+W^-$.}
\end{figure}

The tree-level cross-section $\sigma^0(s,s_1,s_2)$ in (\ref{twonine}) may be
obtained from the Born matrix elements:
\beq
{\cal M}_B = {e^2\over 2s^2_W}~{1\over t}~{\cal M}_1\delta_L + e^2~\left({1\over
s} - \cot\theta_W~g_{eeZ}~{1\over s-m^2_Z}\right)~2~({\cal M}_B - {\cal M}_Z)
\label{twoeleven}
\eeq
where $\delta_L = 1,0$ for $e_{L,R}$ and $g_{eeZ} = \tan\theta_W -
\delta_L~{1\over 2\sin\theta_W\cos\theta_W}$. Close to threshold, one has
${\cal M}_1 \sim 1$ and ${\cal M}_{2,3}\sim\beta$, so the cross-section is
dominated by $t$-channel $\nu$ exchange, which yields an angular distribution
\beq
{d\sigma\over d\theta} = {\alpha^2\over s}~{1\over 4\sin^4\theta_W}~\beta
\left[ 1 + 4\beta\cos\theta~{3\cos^2\theta_W-1\over 4\cos^2\theta_W-1}
+ 0(\beta^2)\right]
\label{twotwelve}
\eeq
and a cross-section
\beq
\sigma = {\pi\alpha^2\over s} \cdot {1\over 4\sin^4\theta_W} \cdot 4\beta +
0(\beta^3)
\label{twothirteen}
\eeq
close to threshold. This yields a sharp threshold rise, but does mean that the
cross-section close to threshold is not very sensitive to the triple-gauge
couplings to be discussed in Section 2.4.

\begin{figure}
\hglue2cm
\epsfig{figure=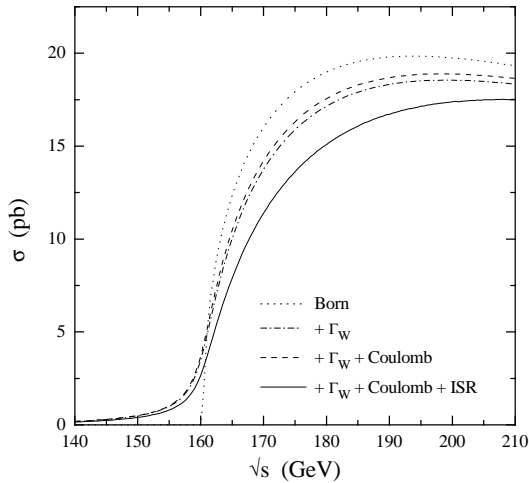,width=7cm}
\caption[]{
Threshold cross section for $e^+e^-\rightarrow W^+W^-$, including finite-width
effects, Coulomb corrections and Initial-State Radiation (ISR).}
\end{figure}

The one-loop electroweak corrections $\delta_{ew}$ in (\ref{twoeight}) are
completely known for on-shell $W^\pm$, but only the leading contributions $\sim
\ln(s/m^2_e)$, $\sqrt{(m_W/\Gamma_W)}$, $m^2_t/m^2_W), \ldots$ are fully known for
off-shell $W^\pm$~\cite{LEP2}. Comparing the ``CC03" cross-section with
calculations of
all the 11 diagrams contributing to $e^+e^-\rightarrow \mu^-\bar\nu u\bar d$
suggests that the residual theoretical error in the cross-section
$\delta\sigma/\sigma \lappeq$ 2\%. As seen in Fig. 15, important r\^oles in
reaching this precision are played by ISR corrections, for which the formalism
in (\ref{onetwentyseven}) can be used, and Coulomb corrections near threshold.
These blow up for stable on-shell particles:
$\delta\sigma /\sigma \sim {\alpha\pi / v_0} : v_0 = \sqrt{1-(4m^2_W/s)}$,
but are cut off in this case by the finite width of the $W^\pm$:
$ {\alpha\pi / v_0} \rightarrow \alpha\pi \sqrt{(m_W/\Gamma_W)}$: the $W^\pm$
decay before they can be bound!~\cite{Coulomb} As seen in Fig. 15, the
Coulomb correction is about 6
\% in the threshold region, corresponding to an error of about 100 MeV in $m_W$ if
it were not included.

The statistical error in measuring the $W^+W^-$ cross-section is
\beq
\Delta \sigma_{W^+W^-} = {\sigma_{W^+W^-}\over\sqrt{N_{W^+W^-}}} =
{\sqrt{\sigma^{W^+W^-}}\over \sqrt{\epsilon_{W^+W^-}{\cal L}}}
\label{twofourteen}
\eeq
where $N_{W^+W^-}$ is the number of events recorded, which depends on the
efficiency $\epsilon_{W^+W^-}$ as well as the accumulated luminosity ${\cal L}$.
The corresponding error in $m_W$ is~\cite{LEP2}

\begin{figure}
\hglue3.5cm
\epsfig{figure=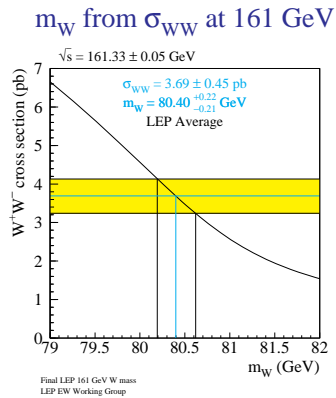,width=4cm}
\caption[]{
Measurement of $m_W$ from the threshold cross-section for $e^+e^-\rightarrow
W^+W^-$ at LEP.}
\end{figure}

\beq
\Delta m_W = \sqrt{\sigma_{W^+W^-}}~\bigg\vert{dm_W\over
d\sigma_{W^+W^-}}\bigg\vert~{1\over\sqrt{\epsilon_{W^+W^-}{\cal L}}}
\label{twofifteen}
\eeq
The first two factors in (\ref{twofifteen}) govern the sensitivity of the
threshold cross-section method of measuring $m_W$, and are minimized around 0.91
GeV/$\sqrt{pb}$ when $E_{cm} \simeq 2m_W$ + 0.05 GeV. On the basis of
this and the
previous world average determination of $m_W$ (not to mention the availability
of superconducting radio-frequency cavities), a centre-of-mass energy
$E_{cm} = 161.33 \pm$ 0.05 GeV was chosen for the threshold LEP 2W run in the
summer of 1996. The accumulated luminosities were about 10 pb$^{-1}$ per
experiment, for which systematic errors associated with $\sigma_{bkgd},
\epsilon_{W^+W^-}$, etc., were relatively unimportant. The resulting measurement
of
$\sigma_{W^+W^-}$ provided the measurement~\cite{LEPEWWG}
\beq
m_W = 80.40^{+ 0.22}_{-0.21}~{\rm GeV}
\label{twosixteen}
\eeq
as seen in Fig. 16.

\subsection{Kinematic Reconstruction of $e^+e^-\rightarrow W^+W^-$}

As already mentioned, this is the way to measure $m_W$ at higher centre-of-mass
$E_{cm} \geq$ 170 GeV. The wholly leptonic final states $(W^+\rightarrow
\ell^+\nu )~(W^-\rightarrow \ell^-\bar \nu )$ are not useful in this respect,
since they have two missing neutrinos and a small branching ratio of about 10 \%.
The semileptonic decays
$(W^\pm\rightarrow \ell^\pm\nu )~(W^\mp\rightarrow \bar qq )$
are much more useful, since they have a larger branching ratio $\sim$ 45 \%, it
is possible to select efficiently a small-background sample, and the final
states can be reconstructed in a constrained fit~\cite{LEP2}. The wholly
hadronic decays
$(W^+\rightarrow \bar qq)~(W^-\rightarrow \bar qq)$ have a similar branching
ratio and one can in pricniple make a more highly-constrained fit, but the
background problems are more severe, and there are the problems of colour
reconnection~\cite{colreconn} (or cross-talk, or exogamy~\cite{EGdiff})
between the hadronic showers of the
$W^\pm$ pair and of distortions induced by Bose-Einstein correlation
effects~\cite{BE} to be discussed shortly.

Figure 17 compares the expected signals and backgrounds in the semileptonic and
purely hadronic final states, and Table 3 compiles the systematic and
statistical errors in the corresponding determinations of $m_W$~\cite{LEP2},
setting aside
possible colour reconnection and Bose-Einstein effects. Most estimates have
suggested that these may be $\leq$ 50 MeV~\cite{LEP2}, but one would like
to be able to
cross-check these estimates\cite{EGWW}. This seems to require a deeper
understanding of
the space-time development of hadronic showers in $e^+e^-$
annihilation~\cite{EGZ}, which we now discuss.

\begin{table}
\begin{center}
\caption{}
\vspace*{0.3cm}
\begin{tabular}{|l|c|c|}  \hline
Source & $W^+W^-\rightarrow q\bar qq\bar q$ & $W^+W^-\rightarrow q\bar q
\ell\nu$
\\
\hline
$E_{beam}$ & 12 & 12 \\
ISR & 10 & 10 \\
fragmentation & 16 & 16 \\
backgrounds & 12 & 6 \\
calibration & 10 & 10 \\
MC statistics & 10 & 10 \\
mass fit & 10 & 10 \\
jet assignment & 5 & - \\
interconnection & ? & - \\ \hline
{\bf Total} & {\bf 31} & {\bf 29} \\
\hline
\end{tabular}
\end{center}
\end{table}

\begin{figure}
\hglue.3cm
\mbox{\epsfig{figure=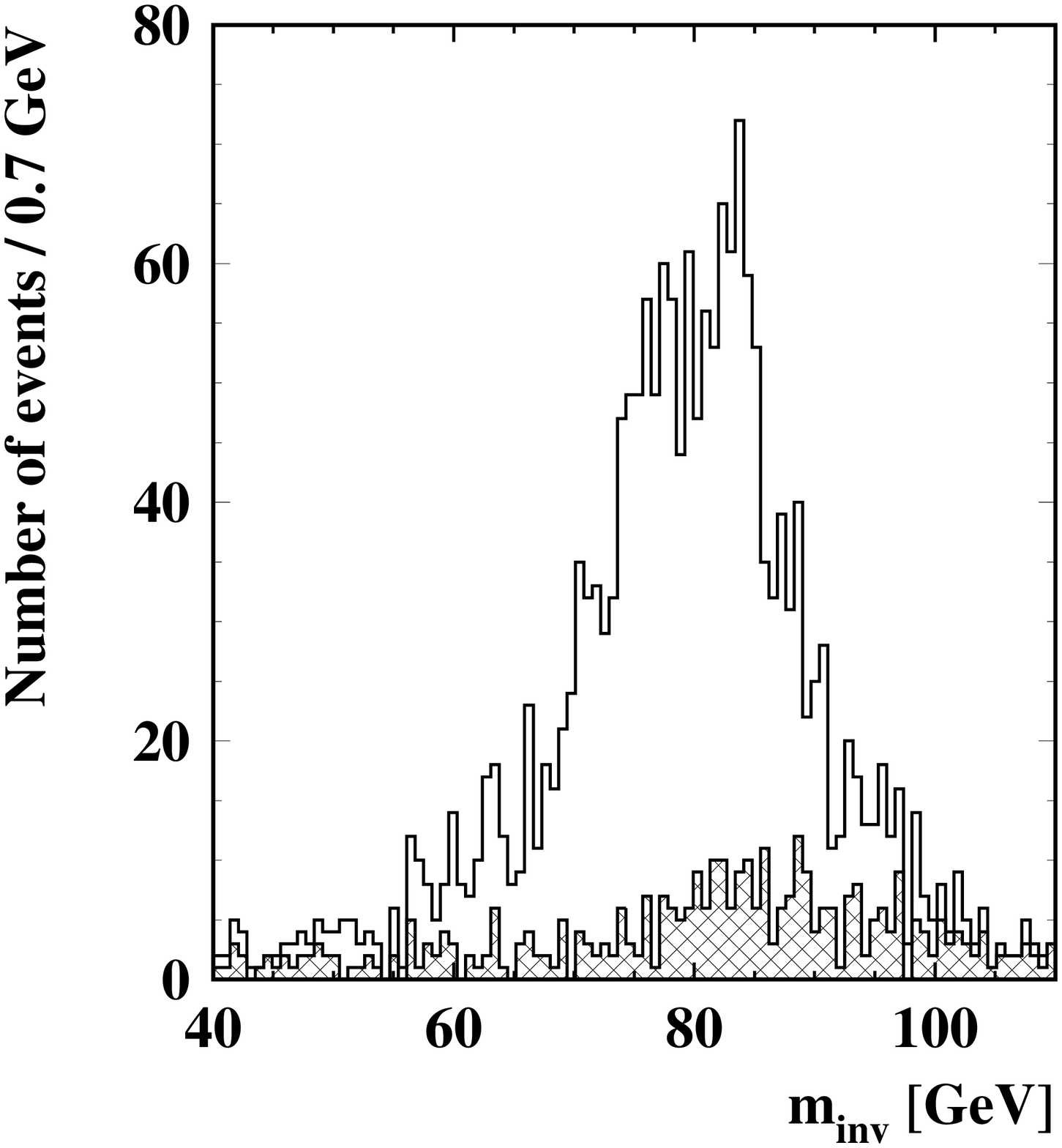,width=5cm}(a)
\epsfig{figure=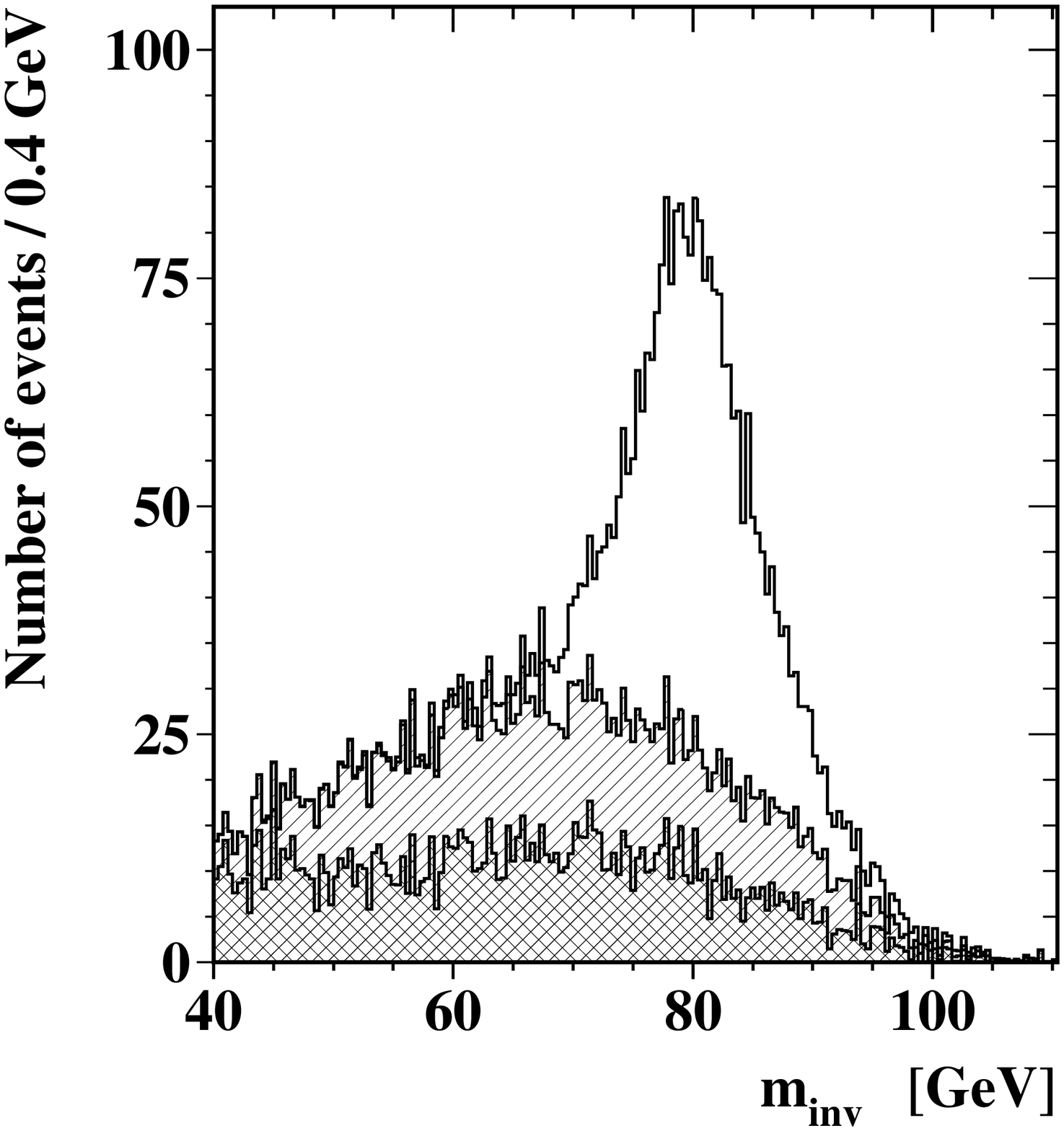,width=5cm}(c)}
\hglue.3cm
\mbox{\epsfig{figure=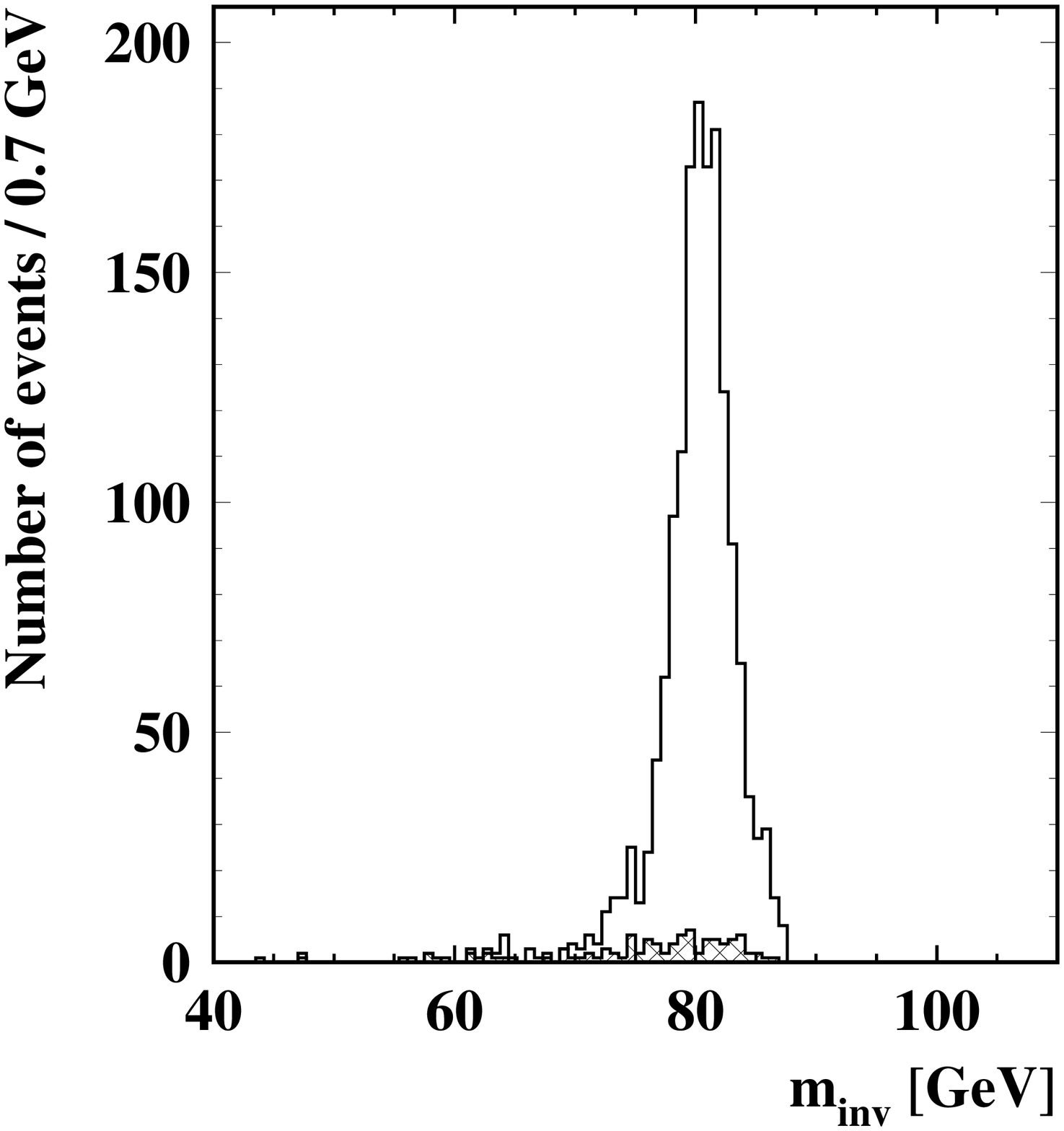,width=5cm}(b)
\epsfig{figure=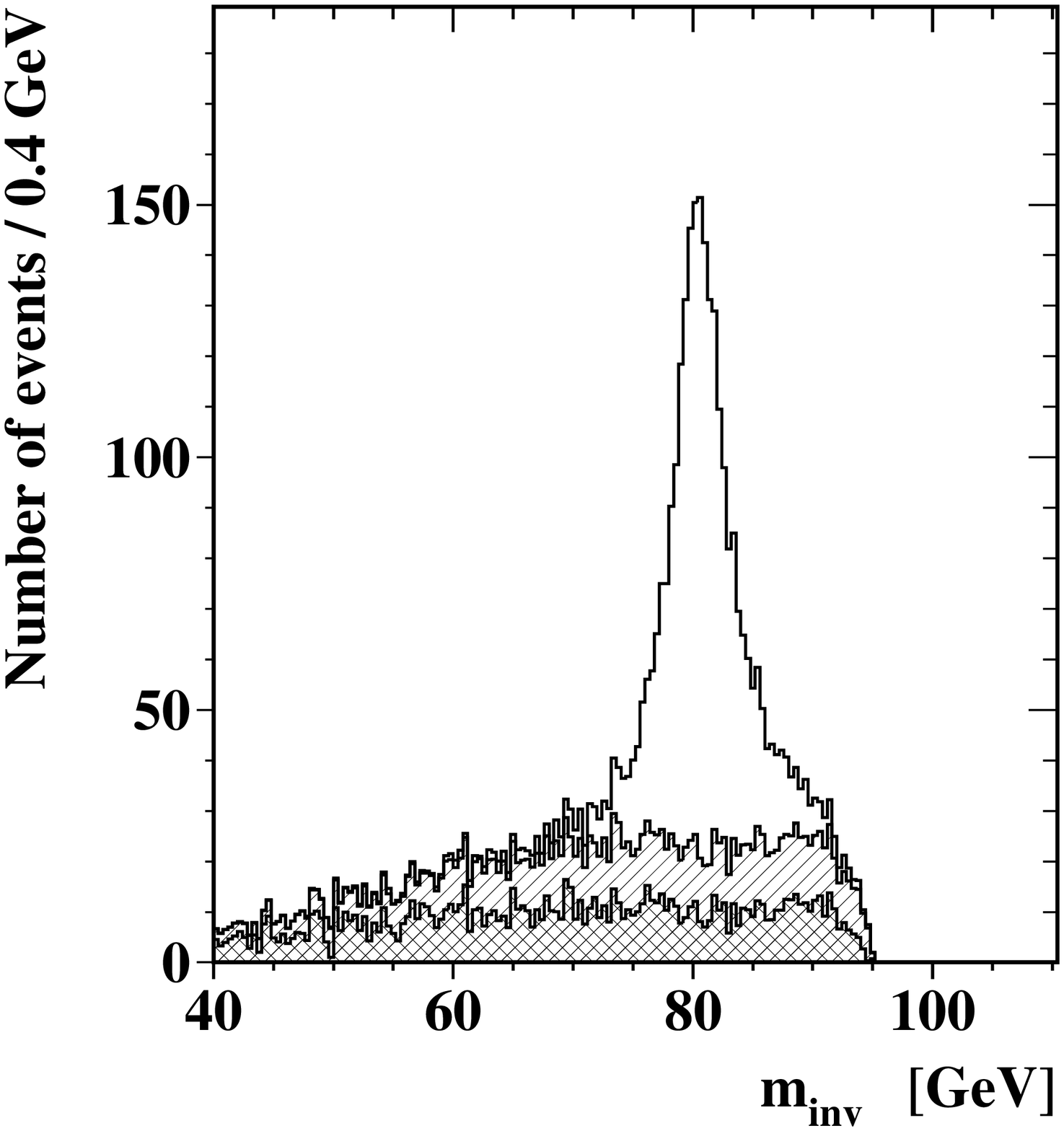,width=5cm}(d)}
\caption[]{
Reconstructed hadronic W mass distributions for (q,b) $W^+W^-\rightarrow\bar
qq\ell\nu$ events and (c,d) $W^+W^-\rightarrow (\bar qq)(\bar qq)$ events, both
before (a,c) and after (b,d) background event deselection.}
\end{figure}

Consider first the process $e^+e^-\rightarrow Z^0\rightarrow \bar qq
\rightarrow$ hadrons~\cite{EGZ}. The initial $Z^0$ decay is to a pair of
``hot" off-shell
partons, which can be regarded as creating a localized ``hot spot" of the
quark-gluon ``parton phase"
of QCD, surrounded by the usual ``cold" hadronic vacuum with condensates
$<0\vert\bar qq\vert 0>$, $<0\vert G^a_{\mu\nu}G^{\mu\nu}_a\vert0>\not= 0$
``frozen" in, as seen in Fig. 18. In the subsequent perturbative parton shower
development, the ``hotter" highly virtual partons decay into ``cooler" partons
closer to mass shell: the ``hot spot" expands and ``cools". Since confinement
forbids the isolation of any parton, whenever a parton separates to $\gappeq$ 1
fm from its nearest-neighbour partons, it should be confined and participate in
hadronization. This takes place at different times in different regions of the
parton shower, and the process continues until all partons ``cool", separate and
hadronize.

\begin{figure}
\hglue2cm
\epsfig{figure=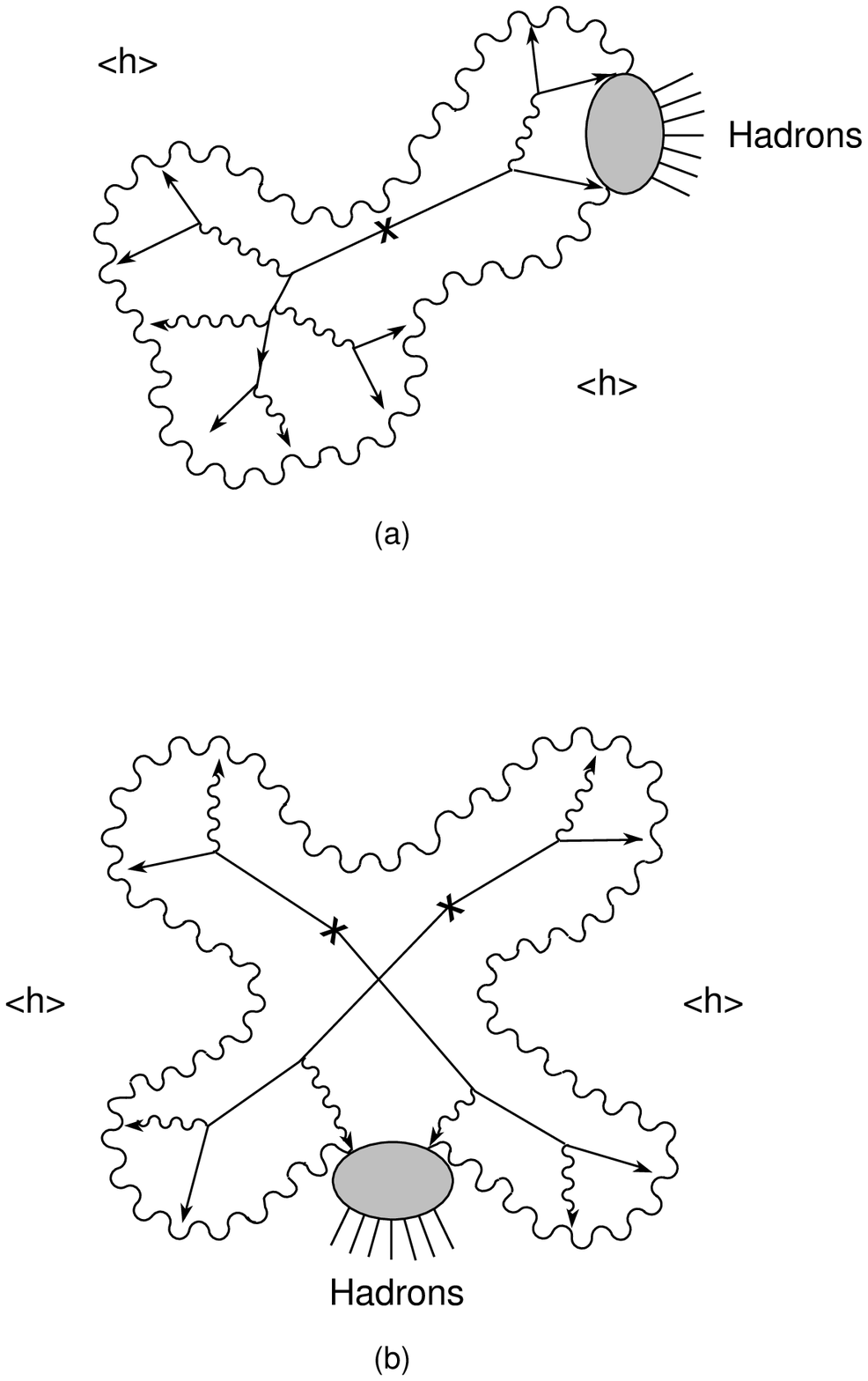,width=8cm}
\caption[]{
Parton shower development and hadronization in (a) $Z^0$ decay and (b) $W^+W^-$
decay is viewed as a ``hot" spot expanding from the decay vertices, marked with
crosses, into the conventional ``cold" hadronic vacuum $<h>$, with clusters
of hadrons
forming whenever a parton gets separated from its nearest neighbour by about 1 fm.}
\end{figure}

The initial stages of this parton shower development can be modelled using a
conventional perturbative QCD Monte Carlo in which spatial locations are also
followed~\cite{EGZ}. The lifetime $t_p$ of a parton obeys $<t_p (x, k^2)> =
\gamma\tau \simeq
E/k^2 = {xQ / 2k^2}$, where $\tau$ is its proper lifetime, $k^2$ is
virtuality, and $x$ its beam energy fraction. Its subsequent chain of decays
into ``cooler" partons lasts a  time after $n$ steps:
\beq
t^{(n)} : <t^{(n)}> = \sum^n_{i=1} <t_i> = {Q\over 2}~\sum^n_{i=1} {x_i\over
k^2_i}~.
\label{twoseventeen}
\eeq
For soft partons in a perturbative QCD cascade, one finds the total time lapse
\beq
<t(x,k^2)> \sim a~{xQ\over 2k^2}~\exp (-b \sqrt{\ln(1/x)})
\label{twoeighteen}
\eeq
corresponding to the familiar ``inside-outside cascade".

This behaviour can be modelled in a perturbative QCD Monte Carlo programme in
which the quantum transport equations are correctly implemented~\cite{G},
including the
time development. This is followed in short-time steps, with branching (and in
principle recombination) processes occurring stochastically at rates with the
means (\ref{twoseventeen}). At each time step, the spatial locations of all the
partons are recorded and the invariant separation of parton pairs measured.
Whenever any parton becomes separated from its nearest neighbour by an amount
$\Delta r \simeq L_c$, where $L_c$ is a critical length parameter, the parton
hadronizes by coalescence with this nearest neighbour, possibly with parton
emission to preserve colour and global quantum numbers. Figure 19 shows the
time developments of parton and hadron numbers and distributions in three
implementations of this non-perturbative model which treat hadronization with
increasing attention to the colour degrees of freedom~\cite{EGZ,EGWW}. They
share the feature
that hadronization occurs gradually over a period ${1\over 2} fm/c \lappeq t
\lappeq 10 fm/c$.

\begin{figure}
\hglue3.5cm
\epsfig{figure=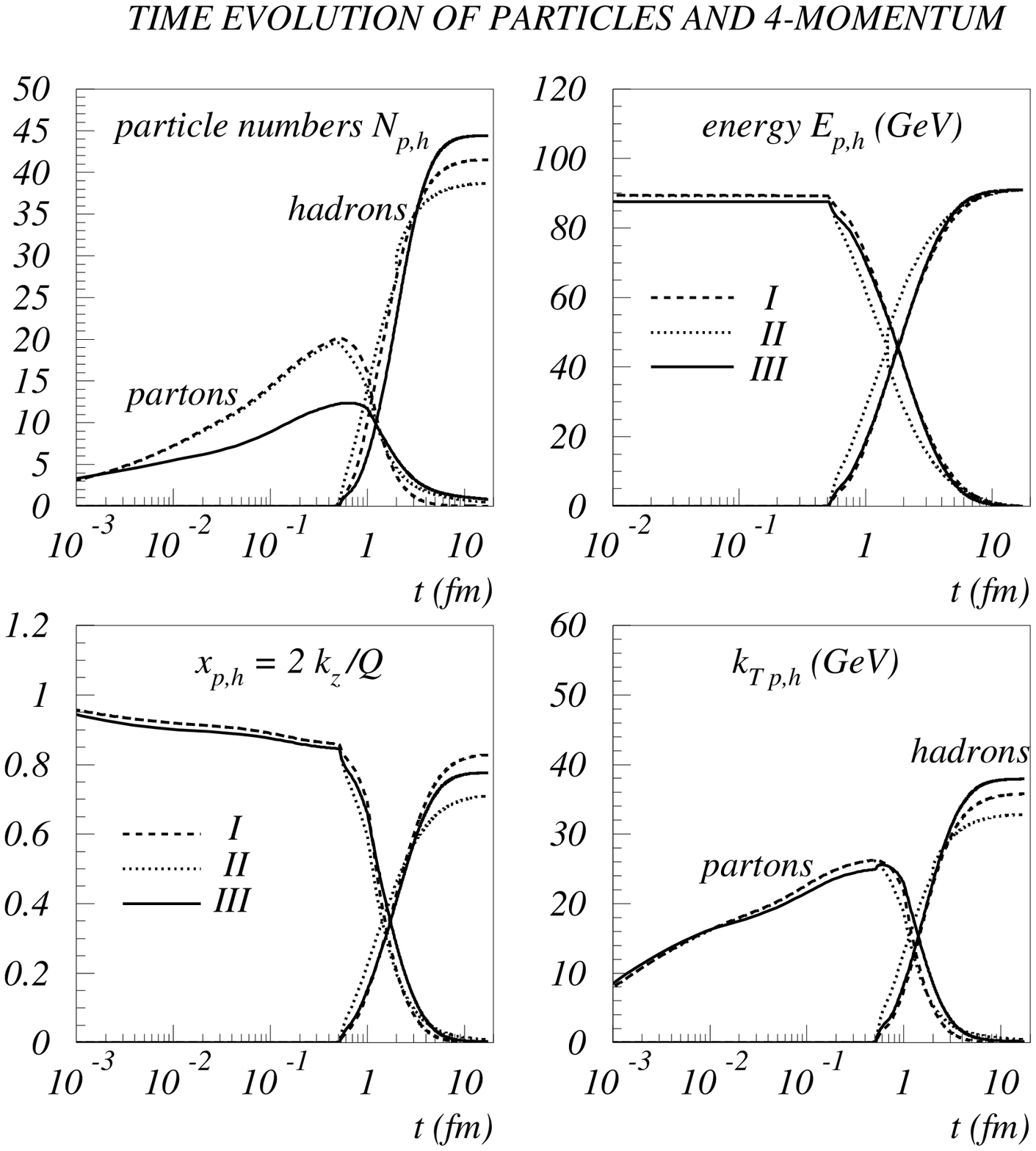,width=6cm}
\caption[]{
Time evolution of parton and hadron distributions in a space-time model for parton
shower development and hadronization, using three different scenarios (I, II, III)
for cluster formation.}
\end{figure}

\begin{figure}
\hglue4cm
\epsfig{figure=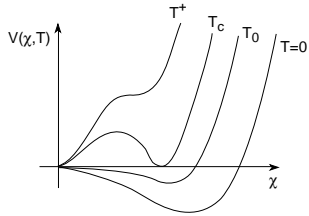,width=4cm}
\caption[]{
Sketch of the temperature dependence of the effective QCD potential, as a
function of a hadronic order parameter $\chi$ related to condensates $<h>$. One
expects similar behaviour for a finite-size QCD system, with $T\rightarrow 1/L$.}
\end{figure}

More specifically, our model for parton-hadron conversion is based on an
effective Lagrangian description of the low-energy ``cool" hadron phase, which
manifests the chiral symmetry of almost massless pions and kaons:
\beq
{\cal L}_{QCD} \rightarrow {\cal L}_{eff} (\pi, K, \ldots)~.
\label{twonineteen}
\eeq
Within this model, the quark-hadron phase transition at finite temperature $T$
and negligible chemical potential can be described using conventional
field-theoretical techniques~\cite{CEO}. One finds a critical temperature
$T_c =
O(\Lambda_{QCD})$ at which the conventional perturbative QCD vacuum and the
``cold" non-perturb\-ative QCD vacuum have equivalent free energies. If one
quantizes the effective theory (\ref{twonineteen}) in a finite volume instead of
at finite temperature, one finds a corresponding transition at a critical size
$L_c = O(1/T_c)$~\cite{EGZ}. The transition is expected to be a tunnelling
phenomenon
analogous to that through the finite-temperature barrier in Fig. 20. The
conversion process is actually stochastic with a probability that is peaked at
sizes $R \gappeq L_c$. Figure 21 shows how the sizes of hadronic clusters in
three different colour implementations of this approach for $e^+e^- \rightarrow
Z^0\rightarrow$ hadrons, together with some resulting hadron
distributions~\cite{EGZ}. The
time evolutions of parton and hadron spectra are shown in Fig. 19. As
seen in Fig. 22, this model reproduces the expected inside-outside cascade
picture.
x
\begin{figure}
\hglue3.5cm
\epsfig{figure=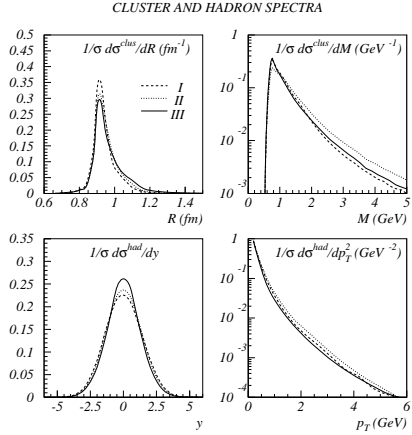,width=5cm}
\caption[]{
Cluster and hadron distributions in the same model as used in Fig. 19.}
\end{figure}

\begin{figure}
\hglue3.5cm
\epsfig{figure=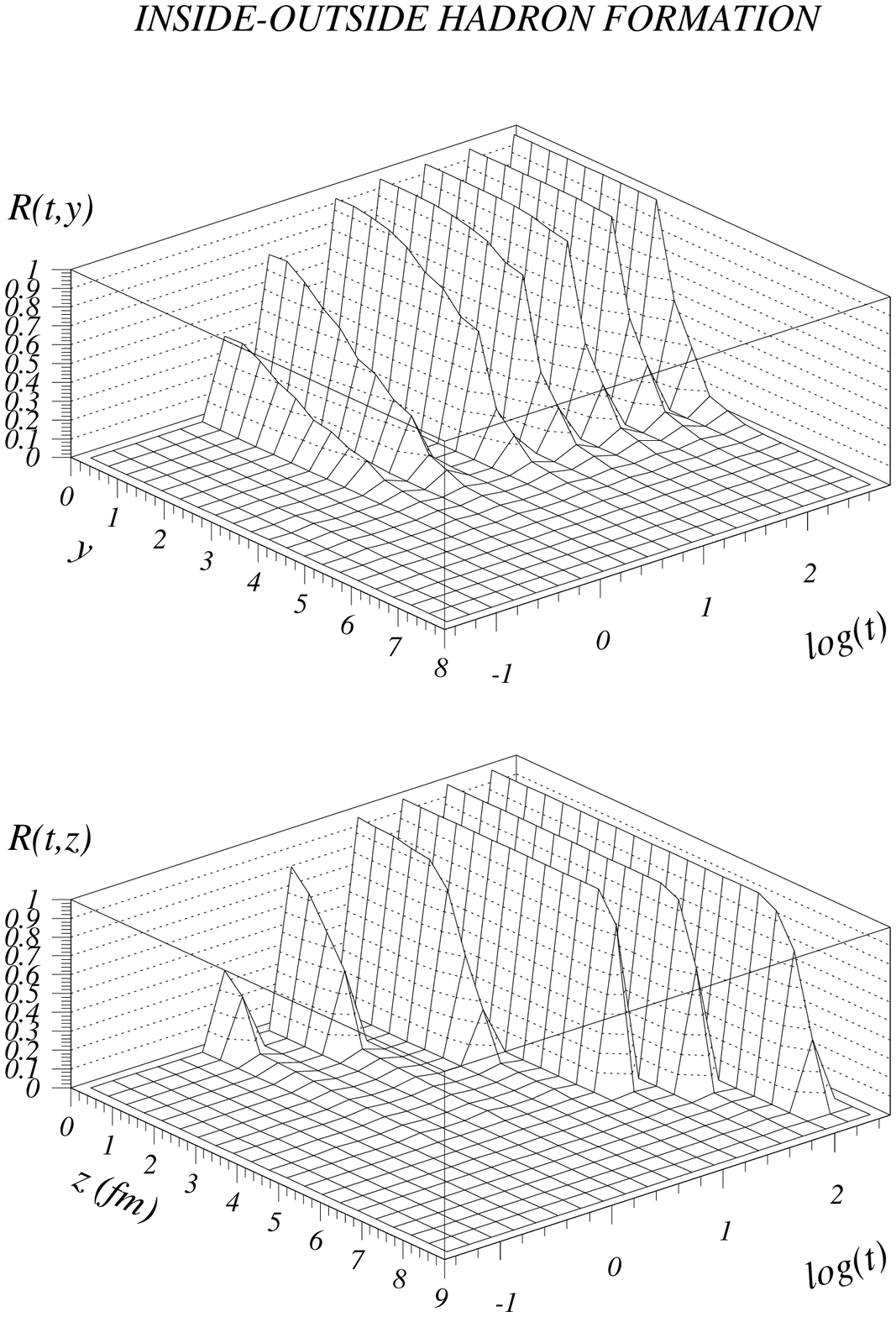,width=5cm}
\caption[]{
Space-time development of the inside-outside cascade in the same model as used in
Figs. 19 and 21.}
\end{figure}

After this introduction, we are now ready to apply this approach to the process
$e^+e^-\rightarrow W^+W^-\rightarrow$ hadrons~\cite{EGWW,EGdiff}.

The first point to notice in this case is that the $W^+W^-\rightarrow (\bar
qq) (\bar qq)$ decays occur almost on top of each other, after times $t^\pm: <
t^*> =
\gamma / \Gamma_W$, separated by a distance $|r^+ - r^-|: <r^\pm > = (\gamma
/\Gamma_W)~(1, \beta^\pm)$. Using the Standard Model value for $\Gamma_W$,
one finds a typical separation $|r^+-r^-| \lappeq 0.1$ fm. Hence, the
$W^+W^-$ decays form a single ``hot spot", not two, as seen in Fig. 18, and are
followed by parton showering that is approximately simultaneous and coincidental.
{\it A priori}, a parton from one $W$ does not ``know" that it should
hadronize with another parton from the same shower in an ``endogamous"
union, and may prefer to hadronize with a parton from the other $W$ in an
``exogamous" union.
 In our approach, each parton chooses its hadronization partner purely on the
basis of its propinquity, not its shower ancestry. This is a non-perturbative
extension of the observation already made some time ago in perturbation
theory that the $W^\pm$ parton showers are not independent~\cite{colreconn}.
However,
perturbative colour reconnection effects are known to be very small:
${\cal O} \left({\alpha_s / \pi}\right)^2 / N_c$.

The fact that the $W^\pm \rightarrow \bar qq$ do not hadronize independently
means that the final-state hadron distributions differ, in general, from
hypothetical ``infinitely-separated" $W^\pm$ pairs.  Indeed, the question
which hadron comes from which $W^\pm$ does not even have a well-defined
answer. In particular, these differences in the final-state hadron
distributions can be expected to have an effect  on $m_W$ which is ${\cal O}
(\Lambda_{QCD})$. We have studied this possible effect in different variants
of our parton-hadron conversion model, using scenarios which differ in their
treatment of the colour book keeping.
We found mass shifts~\cite{EGWW}
\beq
\delta m_W |_{\rm real - hypo} = (-13, +6, +280) ~{\rm MeV}
\label{twotwenty}
\eeq
between the masses extracted using standard jet algorithms in the realistic case
of overlapping $W^\pm$ and a hypothetical situation with $W^\pm$ decays
separately widely in space, which could exceed the statistical and systematic
errors mentioned previously.

In view of the potential gravity of this colour-reconnection effect, it is
desirable to identify possible observational signatures that could give
advance warning of such a phenomenon~\cite{EGdiff}. One possibility is that
the difference
in hadronization could show up directly as a change in the total hadronic
multiplicity:
\beq
< n >_{W^+W^-\rightarrow \bar qq\bar qq} \not= 2 <n>_{(W^\pm \rightarrow \bar
qq)}
\label{twotwentyone}
\eeq
As shown in  Fig. 23, this effect could be substantial, possibly as large as
10 \%.  Such a large effect would also show up
in the mean transverse momenta, as also shown in Fig. 23,
because the  total transverse energies of the initial-state $\bar qq$ jets
have to be shared between fewer hadrons.
Differences may also show up in the longitudinal momentum or rapidity
distributions, as shown in Fig.~24. It is understandable that the effect
should be larger for slower particles, which are more confused about their
origins, since the slower partons overlap more than the faster ones. The
effect extends also to faster partons if the $W^\pm$ decay into  $\bar qq$
dijets that are almost antiparallel, as seen in Fig. 25. The initial
studies of $W^\pm$ final states with data taken during 1996 do not reveal any
gross features of this type~\cite{WWexp}, but the statistics are quite
limited.  It is to
be hoped that the greater statistics to be gathered in 1997 will permit a
more detailed study of this question.

To conclude this section, Fig. 26 shows a compilation~\cite{LEPEWWG} of
present measurements of
$m_W$ from $\bar pp$ colliders (dominated by FNAL~\cite{FNALMW}) and from
LEP2, including both
the threshold and reconstruction methods, and assuming in the latter case that the
hadronic ``exogamy" and Bose-Einstein effects are negligible at the present level
of statistics. Also shown for comparison is the Standard Model prediction, based on
precision electroweak measurements from LEP and the SLC.

\begin{figure}
\hglue2cm
\epsfig{figure=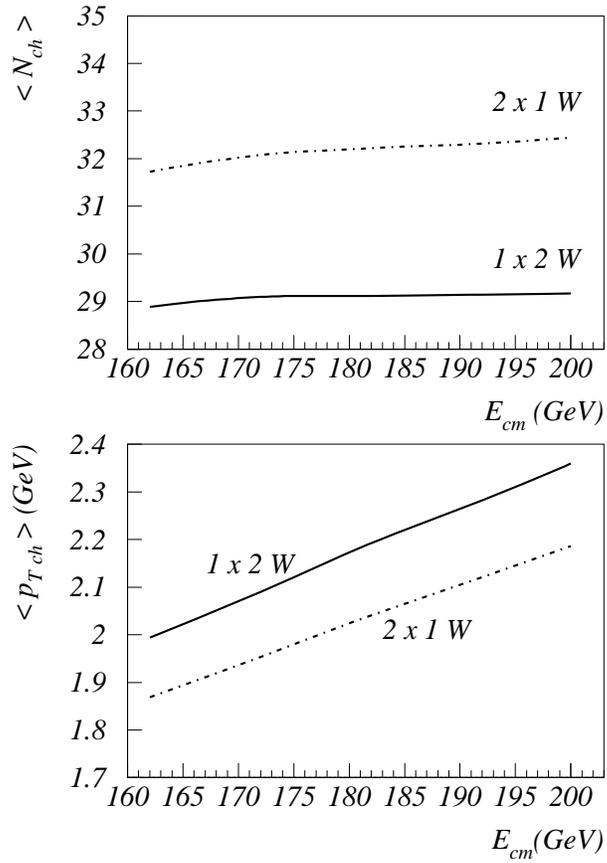,width=8cm}
\caption[]{
Possible differences between charged-hadron multiplicities and transverse momenta in
$e^+e^-\rightarrow W^+W^-\rightarrow (\bar qq)(\bar qq)$ events $(1\times 2W)$ and
independent $W^\pm\rightarrow (\bar qq)$ events $(2\times 1W)$.}
\end{figure}

\begin{figure}
\hglue2cm
\epsfig{figure=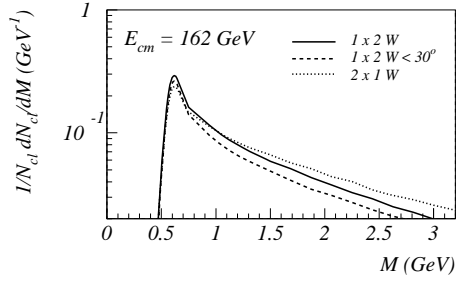,width=6cm}
\caption[]{
Cluster mass distributions in $W^+W^-\rightarrow (\bar qq)(\bar qq)$ (note the
dependence on the angle between the decay dijets) and $W^\pm \rightarrow \bar qq$ events.}
\end{figure}

\begin{figure}
\hglue2cm
\epsfig{figure=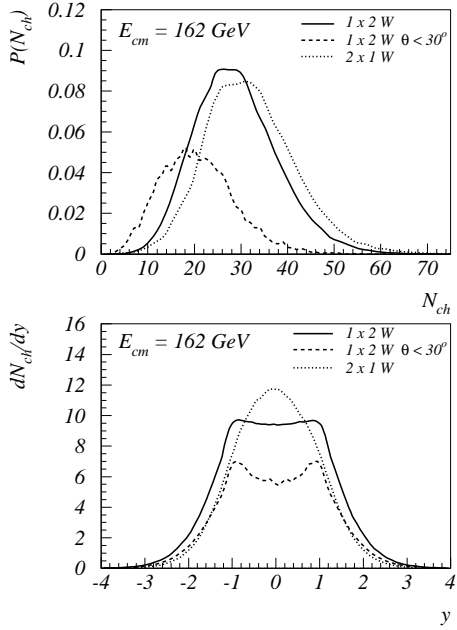,width=6cm}
\caption[]{
Hadron distributions in $W^+W^-\rightarrow (\bar qq)(\bar qq)$ (note again the
angular dependence) and in $W^\pm\rightarrow \bar qq$ events.}
\end{figure}

\begin{figure}
\hglue2cm
\epsfig{figure=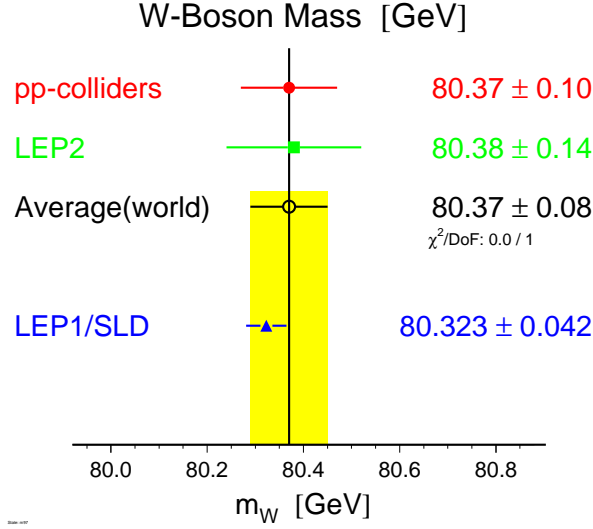,width=8cm}
\caption[]{
Compilation of direct measurements of $m_W$ compared with prediction based on the
Standard Model and precision electroweak measurements from LEP and the SLC.}
\end{figure}

\subsection{TRIPLE-GAUGE COUPLINGS}

The general form of $WWV$ vertex, where $V = \gamma , Z$, may be parametrized
as follows~\cite{LEP2}:
\bea
&&g^\nu_1 V^\mu (W^-_{\mu\nu}W^{+\nu} - W^+_{\mu\nu} W^{-\nu}) + \kappa_v
W^+_\mu W^-_\nu V^{\mu\nu} \nonumber \\
&& + {\lambda_\nu\over m^2_W}~V^{\mu\nu} W^{+\rho}_\nu W^-_{\rho\mu} +
ig^\nu_5
\epsilon_{\mu\nu\rho\sigma} \left((\partial^\rho W^{-\mu}) W^{+\nu} - W^{-\mu}
(\partial^\rho W^{+\nu})\right) V^\sigma \nonumber \\
&& + i g^\nu_4 W^-_\mu W^+_\nu (\partial^\mu V^\nu - \partial^\nu
V^\mu)\nonumber \\
&& - {\tilde\kappa_\nu\over 2} W^-_\mu W^+_\nu
e^{\mu\nu\rho\sigma}\cdot V_{\rho\sigma} - {\tilde\lambda_\nu\over 2m^2_W}
W^-_{\rho\mu} W^{+\mu}_\nu
\epsilon^{\nu\rho\alpha\beta} V_{\alpha\beta}
\label{twotwentytwo}
\eea
Electromagnetic gauge invariance imposes the restrictions $g^\gamma_1 = 1$,
$g^\gamma_5 = 0$ at zero momentum $q^2 = 0$, with possible deviations at
non-zero momentum transfers.
In the Standard Model, one has~\cite{LEP2}
\beq
g^Z_1 = g^\gamma_1 = \kappa_Z = \kappa_\gamma = 1
\label{twotwentythree}
\eeq
whilst the other couplings vanish. More generally, the coupling $g^V_5$
violates C, P but conserves CP, whereas $g^V_4$, $\tilde\kappa_V$, $\tilde
\lambda_V$ violate CP.
For the remainder of this short discussion, we assume C and P invariance, and
make the following convenient multipole parametrization of the remaining
terms~\cite{LEP2}
\bea
Q_W &=& eg^\gamma_1, ~~ \mu_W = {e\over 2m_W}~(g^\gamma_1 + \kappa_\gamma +
\lambda_\gamma) \nonumber \\
q_W &=& -{2\over m^2_W}~(\kappa_\gamma - \lambda_\gamma )~~.
\label{twotwentyfour}
\eea
Furthermore, it is plausible to assume an $SU(2)\times U(1)$
gauge-invariant parametrization,  based on a linear realization, with a single
Higgs doublet~\cite{LEP2}
\bea
&&{ig^\prime\over m^2_W} \alpha_{B\phi} (D_\mu
\Phi)^\dagger~~B^{\mu\nu}(D_\nu\Phi) +
{ig\over m^2_W} \alpha_{W\phi} (D_\mu \Phi)^\dagger
\underline{\tau}\cdot \underline{W}^{\mu\nu} (D_\nu \Phi)\nonumber \\
&&+ {g\over 6m^2_W} \alpha_W \underline{W}^\mu_\nu \cdot (\underline{W}^\nu_\rho
\times
\underline{W}^\rho_\mu)
\label{twotwentyfive}
\eea
Thus we finally boil the fourteen parameters of Eq. (2.22) down to the three
manageable parameters
\bea
\alpha_{W\phi} &=& C^2_W (g^Z_1 - 1) \nonumber \\
\alpha_{W\phi} + \alpha_{B\phi} &=& -{C^2_W\over S^2_W} ~(\kappa_Z - g^Z_1)
= \kappa_\gamma -1 \nonumber \\
\alpha_W &=& \lambda_\gamma = \lambda_Z
\label{twotwentysix}
\eea
which are used in most experimental analyses.

Since any one of these parameters leads to a non-renormalizable growth of
$\sigma (e^+e^- \rightarrow W^+W^-)$ at high energies, they can only be
effective low-energy parameters in a theory that is cut off at some
higher-energy scale $\Lambda_U$. The possible magnitudes of these parameters
depend on this scale $\Lambda_U$:
\beq
\vert\alpha_W\vert \simeq 19\left({m_W\over \Lambda_\nu}\right)^2,~~\vert
\alpha_{W\phi}\vert \simeq 15.5 \left({m_W\over \Lambda_\nu}\right)^2, ~~
\vert\alpha_{B\phi}\vert \simeq 49 \left({m_W\over \Lambda_\nu}\right)^2
\label{twotwentyseven}
\eeq
There are direct bounds on these parameters from the CDF and D0 experiments
at the Fermilab Tevatron collider~\cite{FNAL3V}, as well as from
LEP~2~\cite{LEPEWWG}. As was already
discussed, cross-section measurements close to threshold are relatively
insensitive to the triple-gauge couplings, and one can expect much more
stringent bounds as LEP~2 advances to higher energies and
luminosities~\footnote{It should also be noted that there are indirect
constraints on some combinations of triple-gauge couplings from their virtual
effects on observables at LEP~1.}.

\section{Supersymmetry and LEP~2}
\subsection{(S)Experimental demotivation}
\setcounter{equation}{0}

In the summer of 1995, there was considerable excitation around the
apparently anomalous decay rates $Z\rightarrow \bar bb, \bar cc$: the first
of these  deviated from the prediction of the Standard Model by more than
3 1/2 $\sigma$, and the latter disagreed by about  2 1/2 $\sigma$.
Even if one set $R_c$ to its Standard Model value, $R_b$ still differ from
the Standard Model by 3 $\sigma$~\cite{LEPEWWG95}. There were many warnings
that these
were the most difficult of all the LEP~1 measurements, being dominated by
systematic errors, but this did not stop theorists from speculating that the
appparent discrepencies that might find origins in possible physics beyond
the Standard Model.

In particular, two possible supersymmetric explanations were
proposed~\cite{Rbsusy}. One
involved a light pseudoscalar Higgs A and a very large value of $\tan\beta$,
and the other involved  light charginos
$\chi^\pm$ and a light stop $\tilde t$, and a small value of $\tan\beta$,
which were favoured in some models with an infra-red fixed point. Enthusiasm
for this latter scenario raised hopes that the $\chi^\pm$ and $\tilde t$, might
be detectable in the  LEP~2  energy range.

As you know, this has not yet happened, implying that  the supersymmetric
contribution to $R_b$ cannot be as large as had been speculated.
Already  the non-observation of supersymmetric particles at LEP~1.5, combined
with other phenomenological considerations, indicated that
supersymmetric particles could only provide a small part of the possible
experimental discrepancy in $R_b$~\cite{ELN95}.
We implemented the available experimental constraints on $Z\rightarrow$
nothing visible, $Z\rightarrow \chi\chi^\prime$, the requirements that
$m_{\chi^\pm} \gappeq m_Z/2$ and  $m_h \gappeq $ 40 GeV,
the CLEO constraint  $B(b\rightarrow s\gamma) = $(1 to 4)$ \times 10^{-4}$,
the CDF upper limit on non-Standard Model decays of the top quark, and D0
limits on the stop and neutralino masses.  Finally, we implemented the
LEP~1.5 limit on charginos:  $m_{\chi^\pm} \gappeq$ 65 GeV if $m_{\chi^\pm} -
m_{\chi} \gappeq $ 10 GeV.

We generated  365,000 choices of model parameters with $ 1 < \tan\beta < 5$
and 91,000 more in the restricted range $1 < \tan\beta < 1.5$, with
supersymmetric mass parameters in the range  $0 <  \tilde m < $ 250 GeV.
We found only a handful of models  that made a contribution to $R_b >
0.0010$,  and we found the absolute upper limit~\cite{ELN95}
\beq
\Delta R_b < 0.0017
\label{threeone}
\eeq
We concluded that ``$\ldots$ it may be necessary to review carefully the
calculation and simulation of the Standard Model contributions to $R_b$ and
related measurements.".

New experimental data became available during 1996, and the experimental
discrepancy in $R_b$ and $R_c$ was much diminished to about
$2\sigma$~\cite{LEPEWWG}.
We revisited the calculation of supersymmetric contributions to $R_b$ in the
light of the new exclusions of supersymmetric parameter space from
higher-energy LEP~2 runs~\cite{ELN96}. We implemented the LEP~1, CLEO and
CDF constraints
as before, and implemented the new limits on the lighter stop mass $m_{\tilde
t}$ from  LEP~2 and D0, and new limits on charginos from  LEP~2:
$m_{\chi^\pm} >$ 84 GeV if $m_{\chi^\pm} - (m_{\tilde\nu}$ or $m_\chi)$ are
greater than 3 GeV.  However, we relaxed the previous limits on $m_h$, as a
way of allowing larger values of the heavy second stop mass $m_{\tilde t_2}$.

Following our previous procedure, we generated 484,000 choices of
supersymmetric model parameters, of which 10,000 gave  $\Delta R_b > $
0.0020. After implementation of the constraints listed above, 41,000 models
survived,  of which 210, i.e., 0.04 \% of the original sample, have $\Delta
R_b > $ 0.0010. This emphases the fact that
 large supersymmetric contributions to $R_b$ are  very special~\cite{ELN96}.
Coincidentally, we found a maximum supersymmetric contribution $\Delta R_b
\sim $ 0.0017 again.  However, we emphasized that such a large contribution
to $R_b$ required  choices of parameters that were not attainable in
conventional supergravity models, which could only yield~\cite{ELN96}
\beq
\Delta R_b < 0.0003
\label{threetwo}
\eeq
a truly negligible  contribution to resolving any residual discrepancy in
$R_b$.

The reason for this can be understood from Fig. 27, which shows the
projection on the $(\mu, M_2)$ plane of the ``globular cluster" of surviving
supersymmetric models, compared with the regions of this plane that are
accessible in conventional supergravity models.
This point is also made in Fig. 28, which shows a similar feature in the
$(\theta_t$, $m_{\tilde t_1})$ plane characterizing stop mixing: the
supergravity models miss entirely the
interesting projection of the ``globular cluster".
Moreover, we see in Fig. 29 that the surviving models are very vulnerable to
small improvements in the current experimental exclusion domains in the
$(m_{\tilde t_1}$, $m_\chi)$ plane. Indeed, about a half of the ``globular
cluster" has already been excluded by an improved limit from the OPAL
collaboration~\cite{OPALstop} subsequent to our
analysis~\footnote{The apparent ability of supergravity models to reach into
the ``globular cluster" is only an artefact  of this particular planar
projection: as we saw in the previous figures 27 and 28, supergravity
models are far away in other projections.}.

\begin{figure}
\hglue4cm
\epsfig{figure=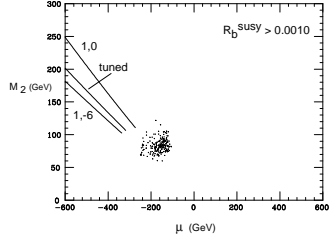,width=4cm}
\caption[]{
The ``globular cluster" of surviving supersymmetric models with $\Delta R_b >$
0.0010, projected on the $(\mu ,M_Z)$ plane and compared with the corresponding
projection for favoured supergravity models.}
\end{figure}

\begin{figure}
\hglue4cm
\epsfig{figure=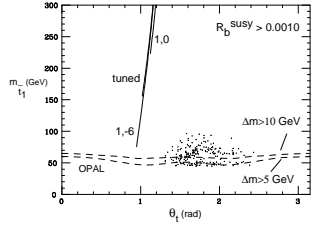,width=4cm}
\caption[]{
As for Fig. 27, but projected on the plane $(\theta_t,m_{\tilde t_1})$ of $\tilde
t$ mixing parameters.}
\end{figure}

\begin{figure}
\hglue4cm
\epsfig{figure=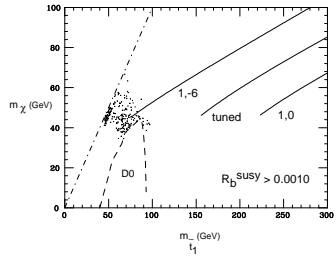,width=4cm}
\caption[]{
As for Fig. 27, but projected on the $(m_{\tilde t_1},m_\chi)$ plane. Note that
none of the favoured supergravity models crosses the ``globular cluster" in the
other projections shown in Figs. 27 and 28.}
\end{figure}

\subsection{Lightest Supersymmetric Particle?}

In many supersymmetric models, this is expected to be stable, and so should
be present in the Universe today as a cosmological relic from the Big Bang.
This stability would be the consequence of the multiplicatively-conserved
quantum number called R parity~\cite{Fayet}, which takes the value +1 for all
conventional particles and -1 for all their supersymmetric partners. The
conservation of R parity  is related to those of baryon $(B)$ and lepton
$(L)$ numbers:
\beq
R = (-1)^{3B+L+2S}
\label{threehtree}
\eeq
where S is the spin, which must be absolutely conserved! Violation of
R conservation is possible if the model violates lepton or baryon number,
either spontaneously or explicitly. If R parity is indeed conserved,  we
have the following three important consequences:

\begin{itemize}
\item Sparticles are always produced in pairs, e.g., $\bar pp \rightarrow
\tilde q \tilde g X$, $e^+e^- \rightarrow \tilde\mu^+\tilde\mu^-$,
\item Heavier sparticles decay into lighter ones, e.g., $\tilde q\rightarrow
q \tilde g$, $\tilde\mu\rightarrow\mu\tilde\gamma$,
\item The lightest supersymmetric particle is stable, because it has no
legal decay mode.
\end{itemize}

However, this is just one of the possibilities for the lightest
supersymmetric particle, as seen in Fig. 30. For example, if R parity is
not conserved, even the lightest supersymmetric particle may decay into
leptons and/or jets, a possibility that has excited renewed interest in the
light of the recent large-$q^2$ events from HERA~\cite{H1,ZEUS}. If R
parity is conserved,
one can ask whether the lightest supersymmetric particle should be neutral,
or whether it could have either electromagnetic and/or strong interactions.
In the latter case, it would interact with ordinary matter and bind to form
anomalous heavy isotopes which are not seen by experiment.
Therefore,  a
stable LSP is presumably neutral with only weak interactions~\cite{EHNOS},
and the
conventional candidate has been the lightest neutralino, as discusssed
below.  In this case, supersymmetry has the ``classic" pure missing-energy
signature. Alternatively, the lightest neutralino might be heavier than some
other supersymmetric particle, such as the gravitino $\tilde G$, in which
case one would have the $\gamma +$ missing-energy signature due to the decay
$\tilde\gamma\rightarrow \gamma\tilde G$.  We will now analyze each of these
possibilities in turn.

\begin{figure}
\hglue2cm
\epsfig{figure=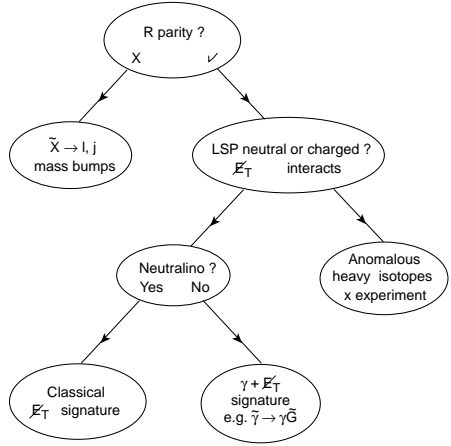,width=6cm}
\caption[]{
Flow chart of possible scenarios for the Lightest Supersymmetric Particle (LSP).}
\end{figure}

\subsection{Constraints on a Stable Neutralino}

The $2\times 2$ chargino and $4\times 4$ neutralino mass matrices of the MSSM
are characterized at the tree level by common parameters: $SU(2)$ and $U(1)$
gaugino masses $M_{2,1}$, the Higgs mixing parameter $\mu$ and the ratio
$\tan\beta\equiv v_1/v_2$ of Higgs vacuum expectation values. The chargino mass
matrix is~\cite{EHNOS,MSSM}
\beq
\left(\matrix{M_2 & g_2{v_2\over \sqrt{2}}\cr
{g_2v_1\over \sqrt{2}} & \mu}\right)
\label{threefour}
\eeq
with two mass eigenstates $\chi^\pm, \chi^{\prime\pm}$, and the neutralino mass
matrix is~\cite{EHNOS,MSSM}
\beq
\left(\matrix{
M_2 & 0 & {-g_2 v_2\over\sqrt{2}} & {g_2 v_1\over\sqrt{2}}\cr\cr
0 & M_1 & {g^\prime v_2\over\sqrt{2}} & {-g^\prime v_1\over\sqrt{2}}\cr\cr
{-g_2v_2\over\sqrt{2}} & {-g^\prime v_2\over\sqrt{2}} & 0 & \mu \cr\cr
{g_2 v_1\over\sqrt{2}} & {-g^\prime v_1\over\sqrt{2}} & \mu & 0}\right)
\label{threefive}
\eeq
with four mass eigenstates $\chi_i$. Gaugino mass universality $M_3 = M_2 = M_1
= m_{1/2}$ is often assumed at the grand unification scale, so that
\beq
{M_2\over M_1}
\simeq {\alpha_2\over\alpha_1} = {8\over 3\sin^2\theta_W}
\label{threesix}
\eeq
at the electroweak scale. Other supersymmetric model parameters enter when one
discusses the production and decays of charginos and neutralinos, their
annihilation in the early Universe and their scattering off nuclei. These
include the scalar masses
\beq
\tilde m^2_i = m^2_{0_i} + C_i m^2_{1/2} + {\cal O} (m^2_Z)
\label{threeseven}
\eeq
where the coefficients $C_i$ are calculable using the
renormalization group, and the $m_{0_i}$
are soft supersymmetry-breaking scalar masses, that may (or
may not) be universal at the GUT scale, as well as the physical Higgs boson
masses. The latter may be characterized by one additional mass parameter $m_A$
at the tree level, but are subject to important radiative corrections that
depend in particular on $m_t$ and the $m_{\tilde t_i}$, as discussed in Section
1.3~\cite{delMH}. There are also interesting radiative corrections to the
chargino and
neutralino masses~\cite{delino}, as we  discuss later, which may be
neglected in a first approximation.

Important constraints on chargino and neutralino masses are imposed by searches
for $e^+e^- \rightarrow \chi^+\chi^-$ and $\chi_i\chi_j$ at LEP~1 (in $Z$ decays)
and at higher energies. Neither the LEP~1 nor the LEP~1.5 data by themselves
provided an absolute lower limit on the possible mass of the lightest neutralino
$\chi$. However, as seen in Fig. 31, the LEP~1.5 data filled in some wedges of
the $(\mu , M_2)$ plane left uncovered by the LEP~1 data for $\mu < 0$ and
$\tan\beta \lappeq 2$, at least for large $m_{\tilde\nu} \gappeq$ 200 GeV,
enabling the absolute lower limit $m_\chi \geq$ 12.8 GeV to be
established~\cite{ALEPHchi}, as
seen in Fig. 32. However, there were loopholes in this first purely
experimental analysis. One appeared for $\tan\beta\sim\sqrt{2}$ and $m_0\sim$ 60
GeV, when the decay $\chi^\pm\rightarrow\tilde\nu + e^\pm$ provided a soft
lepton that was difficult to detect, diminishing the $\chi^\pm$ detection
efficiency. There was also a very small loophole for $1 < \tan\beta < 1.02$,
even at large $m_{\tilde\nu}$, that could probably be filled in by other LEP
data.

\begin{figure}
\hglue2cm
\epsfig{figure=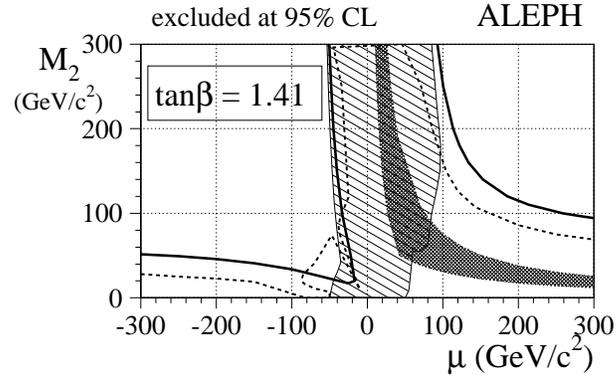,width=8cm}
\caption[]{
Interplay of gaugino constraints from LEP~1 (dashed line) and LEP~1.5 (thick
solid line: $\chi^+\chi^-$, thin solid line $\chi\chi^\prime$). Notice their
complementarity in the $\mu < 0$ quadrant.}
\end{figure}

\begin{figure}
\hglue4cm
\epsfig{figure=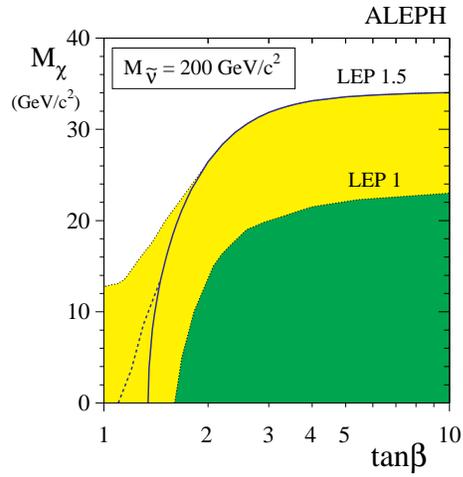,width=6cm}
\caption[]{
Experimental lower limit on $m_\chi$, assuming a large sneutrino mass
$m_{\tilde\nu}$ = 200 GeV and ignoring a small loophole for $1.00 < \tan\beta <
1.02$. There is another, larger, loophole if $m_{\tilde\nu}\simeq$ 60 GeV and
$\tan\beta\simeq\sqrt{2}$.}
\end{figure}

We pointed out~\cite{EFOS1} that these loopholes could be
plugged, and the lower limit on $m_\chi$ strengthened, by including other
experimental constraints, as well as considerations based on cosmology,
astrophysics and dynamical electroweak symmetry breaking, at the price of some
assumptions of universality of the $m_{0_i}$. Neutrino counting at the $Z$ peak
tells us that $m_{\tilde\nu} >$ 43.1 GeV if all three sneutrino
species are degenerate, and  LEP~1.5 searches for charged sleptons already
told us that $m_{\tilde\ell^\pm} \gappeq$ 45 to 60 GeV, although these did not
exclude $\chi^\pm\rightarrow\tilde\nu$ + soft $\ell^\pm$ decay. However, upper
limits on $e^+e^-\rightarrow\gamma$ + nothing from lower-energy accelerators,
notably TRISTAN at KEK~\cite{AMY}, excluded a domain of the $(m_0,m_{1/2})$
plane that ruled
$m_\chi = 0$ out in the region $\tan\beta \sim\sqrt{2}, m_0\sim$ 60 GeV. Further
constraints on $(m_0,m_{1/2})$ were obtained  if one further assumed that the
cosmological relic density of neutralinos
$\rho_\chi =
\Omega_\chi\rho_c$, where $\rho_c$ is the critical cosmological density, fell
into the range
\beq
0.1 \leq \Omega_\chi h^2 \leq 0.3
\label{threeeight}
\eeq
where $h$ encodes the current Hubble expansion rate: $H_0$ = 100 h km s$^{-1}
M^{-1}_{pc}$. The upper limit  (\ref{threeeight}) is an absolute requirement
if $\Omega_{total} \leq 1$ and the age of the Universe $t_0 \geq$   12 Gy,
whereas the lower limit in (\ref{threeeight}) is merely a preference for the
neutralino  to have a relic density large enough to be of astrophysical
relevance. Under these assumptions, we found that the limits $m_{1/2}\rightarrow
0$, $\mu\rightarrow 0$ could be excluded, and we found the lower
limit~\cite{EFOS1}
\beq
m_\chi \gappeq 21.4~{\rm GeV}
\label{threenine}
\eeq
occurring when $\tan\beta\simeq$ 1.6, with $m_\chi$ necessarily much larger for
generic values of $\tan\beta$, as seen in Fig. 33.

\begin{figure}
\hglue2cm
\epsfig{figure=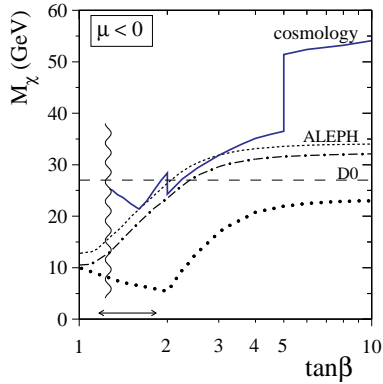,width=5cm}
\caption[]{
Phenomenological lower limits on $m_\chi$ based on LEP~1.5 data, for arbitrary
$m_0$, including the AMY result (dotted line), inferred from the $D\phi$ gluino
search assuming universal gaugino masses (dashed line), assuming scalar-mass
universality (dot-dashed line), and applying the cosmological constraint (3.8)
(solid line).}
\end{figure}

\begin{figure}
\hglue3cm
\epsfig{figure=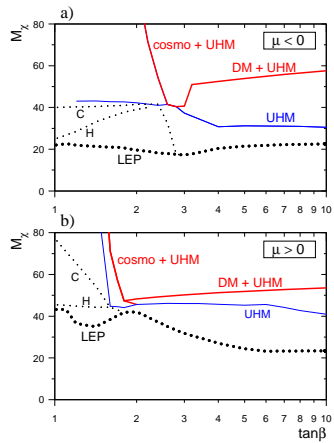,width=5cm}
\caption[]{
Lower limits on $m_\chi$ based on data from LEP~1, 1.5 and 2. The dotted line makes
no appeal to extra theoretical assumptions. Lines labelled UHM assume universal
scalar masses also for Higgs bosons. The branches labelled ``cosmo" and ``DM" assume
the upper and lower limits in Eq. (3.8), respectively. The lines labelled
C and H are explained in~\cite{EFOS2}.}
\end{figure}

We have recently updated this analysis, including the latest chargino and
neutralino limits from the higher-energy LEP~2W and LEP~2 runs, and exploring
systematically the consequences of extending the universal scalar-mass
assumption to more sfermion species~\cite{EFOS2}. The primary experimental
searches are those for $e^+e^-\rightarrow\chi^+\chi^-$, which enforce
\beq
m_{\chi^\pm} \gappeq 85~{\rm GeV} \times f(\mu,m_{\tilde\nu},\tan\beta)
\label{threeten}
\eeq
and for $e^+e^-\rightarrow\tilde e^+\tilde e^-$, which impose
\beq
m_{\tilde e} \gappeq 70~{\rm GeV} \times f(\mu,m_{1/2},\tan\beta)
\label{threeeleven}
\eeq
Assuming input slepton mass universality: $m_{\tilde e_R} = m_{\tilde e_L} =
m_{\tilde\nu_L}$, the low-$m_0$ loophole is largely filled in, and the
$\tan\beta <$ 1.02 loophole completely disappears. Also of great importance are
the LEP searches for $e^+e^-\rightarrow h+Z$. If one extends the universality
assumption to the squarks, these provide important lower limits on
$(m_{0_1},m_{1/2})$, in order that the $h$ boson be heavy enough to have escaped
detection. These searches also tend to fill in the low-$m_0$ hole. Further
strengthening of the lower limit on $m_\chi$ is found if one extends the
scalar-mass universality assumption to the Higgs bosons. This assumption fixes
$\mu$ and $m_A$ as functions of the other parameters, refining the direct LEP
search limits on charginos and sleptons, and enabling the search for
$e^+e^-\rightarrow
h+A$ to come into play. Finally, dramatic strengthening of the lower limit on
$m_\chi$ is found at low $\tan\beta$ if one imposes the cosmological upper limit
on $\Omega_\chi h^2$ (\ref{threeeight}). Combining these considerations, as
seen in Fig. 34, we find~\cite{EFOS2}

\beq
m_\chi \gappeq 40~{\rm GeV}
\label{threetwelve}
\eeq
whatever the value of $\tan\beta$, and also lower limits  $\tan\beta \gappeq$
1.7 if $\mu < 0$ and $\tan\beta \gappeq$ 1.4 if $\mu > 0$.

The latter results do not apply if one relaxes the scalar-mass universality
assumption, enabling one to consider the possibility that the lightest
neutralino is mainly a higgsino. This must weigh less than 80 GeV, otherwise it
would annihilate into $W^+W^-$ pairs in the early Universe, leaving an
uninterestingly low relic density. On the other hand, in this region of
parameter space LEP~2 searches would have observed $\chi^+\chi^-$ production if
$m_{\chi^\pm} \lappeq$ 80 GeV and $m_{\chi^\pm} - m_\chi \gappeq$ 5 GeV. We
infer that there remains only a narrow window 80 GeV $\gappeq m_\chi \gappeq$ 75
GeV for higgsino-like dark matter~\cite{EFOS2}. Just where this region lies
in the
conventional $(\mu, M_2)$ plane depends sensitively on  quantum corrections to
$m_{\chi^\pm}$ and $m_\chi$~\cite{delino}, and hence on $\Omega_\chi
h^2$~\cite{delomega}, which is very
sensitive to $m_{\chi^\pm} - m_\chi$. The interplay of theoretical, experimental
and cosmological considerations in the presence of these quantum corrections
merits deeper study.
x\subsection{R-Conserving Neutralino Decay?}

Decay of the lightest neutralino into a lighter sparticle, such as the gravitino
$\tilde G$, is a generic possibility in no-scale supergravity
models~\cite{EEN} in which \beq
m_{\tilde G} = 0\left( \left({m_W\over m_P}\right)^p\right)~~m_P~:~~p > 1
\label{threethirteen}
\eeq
is achievable, and in gauge-mediated ``messenger" models of supersymmetry
breaking~\cite{messenger}. The radiative decay
$\chi\rightarrow\gamma\tilde G$ would decay inside
the experimental apparatus if $m_{\tilde G}$ is sufficiently small, producing a
$\gamma\gamma + E\llap{$/$}_T$ signature from sparticle-pair production.

Interest in such models has been resuscitated by the (in)famous $\bar
pp\rightarrow e^+e^-\gamma\gamma + E\llap{$/$}_T + X$ event reported by the CDF
collaboration~\cite{CDFevent}. Among the possible interpretations are
$\chi^+\chi^-$ pair production (with $m_{\chi^\pm}
\lappeq$ 150 GeV to get a large enough rate), followed by $\chi^\pm\rightarrow
e^\pm\nu \gamma \tilde G$ decay, and $\tilde e^+\tilde e^-$ production (with
$m_{\tilde e} \gappeq$ 100 GeV to get a large enough rate), followed by $\tilde
e^\pm \rightarrow e^\pm\gamma\tilde G$ decay~\cite{interpretations}.

These interpretations are significantly constrained by LEP~2 searches for
$e^+e^-\rightarrow (\chi\rightarrow\gamma\tilde G)(\chi\rightarrow\gamma\tilde
G)$~\cite{ELNgrav}. As seen in Fig. 35a, preliminary LEP~2W results already
excluded a
considerable fraction of the parameter space in the $\chi^+\chi^-$
interpretation of the CDF event, depending on the value of $m_{\tilde \ell^\pm}$
assumed. A large fraction of the parameter space in the
$\tilde\ell^+\tilde\ell^-$ interpretation has also been excluded, as seen in
Fig. 35b, and the fate of this option may be decided by future higher-energy
LEP runs. Perhaps last week's gravity-mediated models of supersymmetry breaking,
with a heavy gravitino $m_{\tilde G}  > m_\chi$, were not so bad after all?

\begin{figure}
\mbox{\epsfig{figure=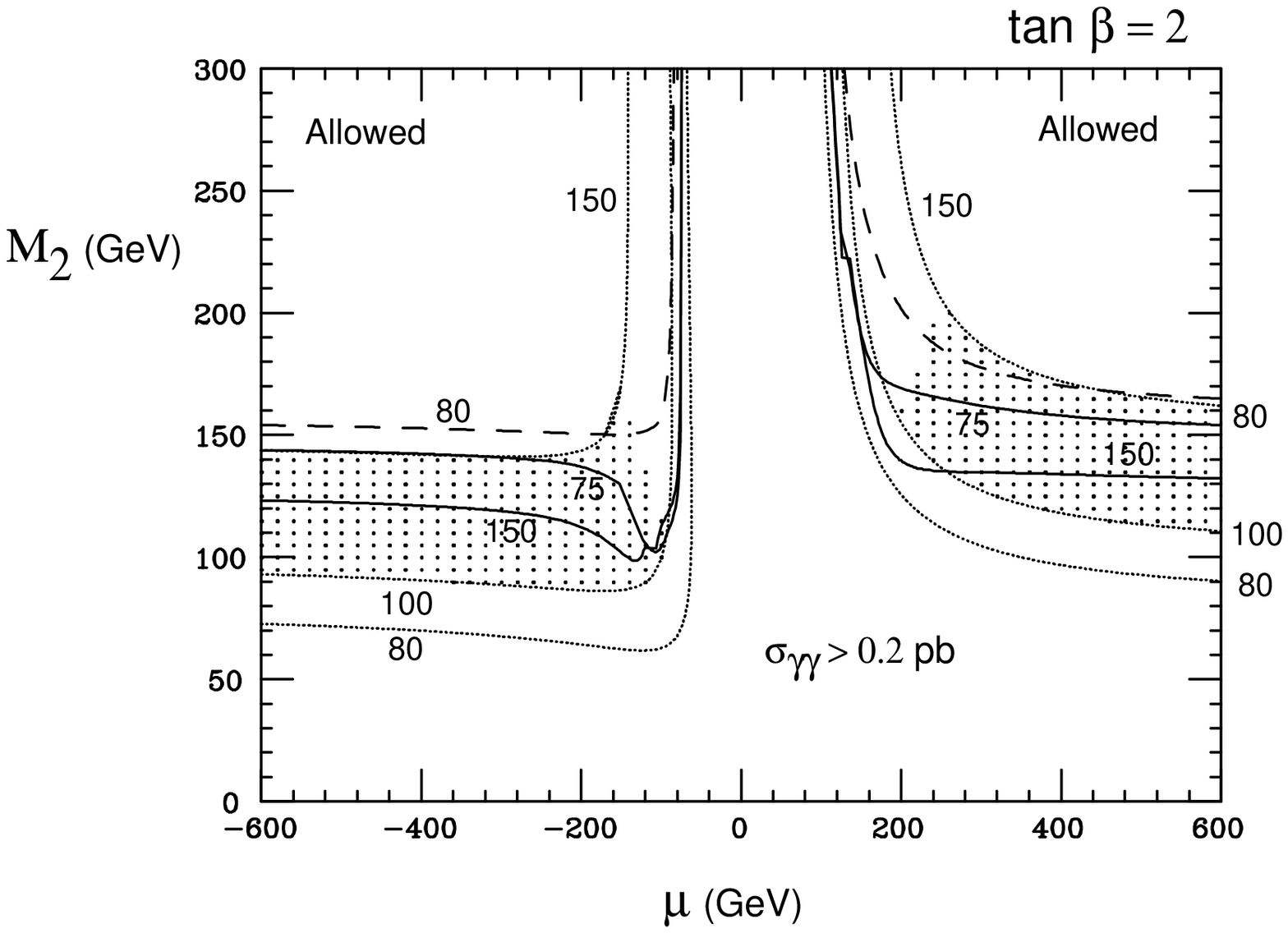,width=5cm}(a)
\epsfig{figure=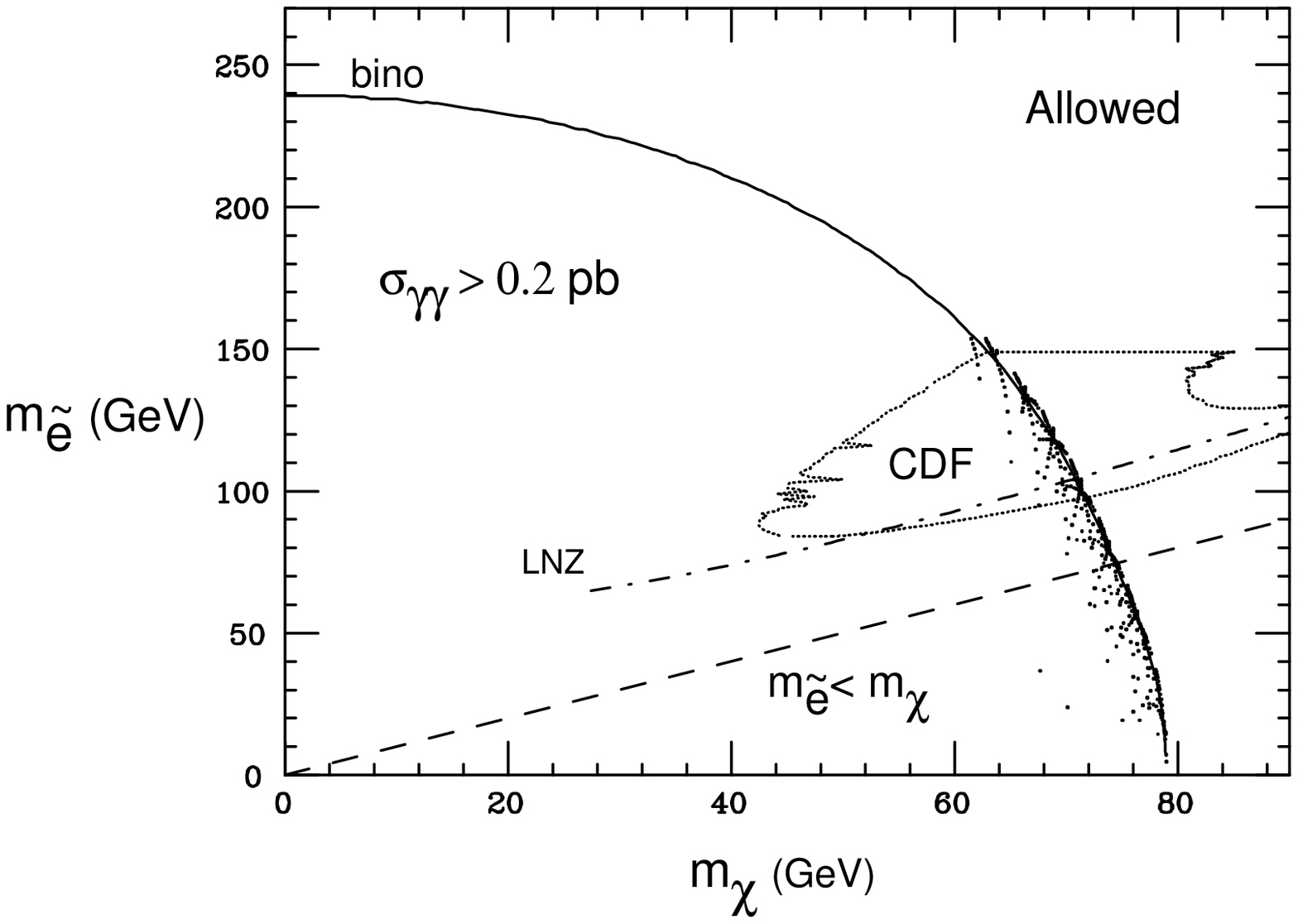,width=5cm}(b)}
\caption[]{
Parameter spaces of favoured light-gravitino models postulating (a) chargino
(dotted regions) and (b) selectron (CDF ``shark's tooth") decay to explain the CDF
$e^+e^-\gamma\gamma E\llap{$/$}_T$ event, compared with preliminary LEP~2
constraints, (a) for different values of $m_{\tilde \ell}$, and (b) for different
values of $m_\chi$ .}
\end{figure}

\subsection{R Violation ?}

In addition to the Yukawa superpotential terms which give masses to the quarks
and leptons, and to the Higgs mixing $\mu$ term, there are other superpotential
terms allowed by the gauge symmetries of the MSSM~\cite{Rviolation}:
\beq
\epsilon_i HL_i + \lambda^{ijk}_1 L_i L_j E^c_k +
\lambda^\prime_{ijk} L_i Q_j D^c_k +
\lambda^{\prime \prime}_{ijk} U^c_i D^c_j D^c_k
\label{threefourteen}
\eeq
each of which violates lepton and/or baryon number, and hence $R$ parity. The
$HL_i$ mixing terms may be removed
by a change in field basis, and the last two terms cannot be present
simultaneously, as they would cause rapid baryon decay. Any of the three
trilinear interactions in (\ref{threefourteen}) would provide dramatic new decay
signatures for sparticles: $\tilde X \rightarrow \ell\ell$, $\ell q$ or $qq$.
Indeed, $R$-violating models might even be easier to spot than the conventional
$R$-conserving scenario. For example, the process $e^+e^-\rightarrow\chi\chi$
becomes observable at LEP because the lightest neutralino $\chi$ has visible
decay modes.

$R$-violating models have been unfashionable for several reasons. One is that,
since $\chi$ is unstable, it is no longer a good candidate for cold dark matter
in the Universe. Another is that $R$-violating couplings would be strongly
constrained by cosmology if the observed baryon asymmetry has a primordial
origin~\cite{CDEO,Fischler}. This is because the $B$- and/or $L$-violating
interactions in (\ref{threefourteen}), in conjunction with the
$(B+L)$-violating non-perturbative interactions in the Standard Model, would
eradicate any primordial $B$ or $L$ asymmetry. Persistence of such an asymmetry
would provide upper limits~\cite{DE}
\beq
\left\vert{\epsilon_i\over\mu}\right\vert \lappeq 3\times
10^{-6}~,~~\vert\lambda\vert\lappeq 10^{-7}
\label{threefifteen}
\eeq
for generic flavour structures of $R$-violating couplings. However, it has been
proposed that one might be able to evade these constraints by playing
flavour-symmetry games, and the cosmological baryon asymmetry might have
originated at the electroweak phase transition, obviating the bounds
(\ref{threefifteen})~\cite{CDEO,DR}. Another potential difficulty for
$R$-violating models is
the observed flavour conservation in neutral interactions, which is natural in
the Standard Model but not in its known $R$-violating extensions.

Some phenomenological interest in $R$-violating models was sparked by the ALEPH
report of a possible signal in $e^+e^-\rightarrow$ 4 jets~\cite{ALEPH4jet},
interpretable as
associated production of a pair of particles decaying into dijet pairs. The
least implausible model for this observation that I know postulates the reaction
$e^+e^-\rightarrow \tilde e^\pm_L \tilde e^\pm_R$, with both $\tilde e^\pm$
decaying into $\bar qq$ via $R$-violating couplings~\cite{CGLW}. This
could be compatible
with the ALEPH data if $m_{\tilde e_R} \sim$ 48 GeV, $m_{\tilde e_L} \sim$ 58
GeV, and one chooses carefully other parameters of the model so as to avoid
large cross-sections for $e^+e^-\rightarrow\tilde\nu_e\tilde{\bar\nu_e}$ and/or
$\chi\chi$ production. However, since the ALEPH signal has not been confirmed by
the other LEP experiments~\cite{fourjetWG}, interest in it has waned.

On the other hand, interest in $R$-violating models has waxed enormously after
the report by the H1~\cite{H1} and ZEUS~\cite{ZEUS} collaborations of a
possible
excess of $e^+p\rightarrow e^+qX$ events at large $Q^2$. Superficially,
the $Q^2$ distribution resembles more what one would expect from a
direct-channel spin-0
resonance than a contact interaction~\cite{AEGLM}, and this is also
suggested by the
 $x$ measurements of the H1 collaboration~\cite{H1}, which are more precise
than
those of ZEUS~\cite{ZEUS}. However, production of a non-supersymmetric
leptoquark may be
difficult to reconcile with limits established by the CDF
collaboration~\cite{CDFLQ}, since
its branching ratio into $e^+q$ could only with difficulty be much less than
unity.

Within supersymmetry the natural interpretation~\cite{H,KK,CR,AEGLM,D}
would use the
$\lambda^\prime_{ijk}$ interaction in (\ref{threefourteen}) to produce a
charge-2/3 squark: $e^+_{d_{k_R}} \rightarrow \tilde u_{L_j}$. Production of
$\tilde u_L/\tilde c_L/\tilde t$ off a valence $d$ quark would require
a coupling $\vert\lambda^\prime_{1j1}\vert \sim 1/25$, whilst production
off a
sea $s$ or $b$ quark would require $\vert\lambda^\prime_{1jk}\vert\sim 1/3$.
Production of the $\tilde u_L$ would conflict~\cite{AEGLM} with the
following upper bound from  nuclear $\beta\beta$ decay~\cite{KKG}:
\beq
|\lambda^\prime_{111}| < 7\times 10^{-3} \left({m_{\tilde q}\over 200~{\rm
GeV}}\right)^2~~\left({m_{\tilde g}\over 1~{\rm TeV}}\right)^{1/2}
\label{threesixteen}
\eeq
whilst production of the $\tilde c_L$ off the $d$ quark is barely compatible
with upper limits from searches for $K\rightarrow\pi\bar\nu\nu$
decay~\cite{AEGLM}. On the
other hand, $\tilde t$ production off either $d$ or $s$ quarks seems to be
compatible with all known constraints~\cite{ELS}.

One wants the branching ratio $B(\tilde q \rightarrow e^+q)$ not to be very
small -- in order that HI and ZEUS have a signal to see -- and not too close to
unity -- otherwise CDF sould have seen $\tilde q \bar{\tilde q}$ pair
production~\cite{CDFLQ}.
In the case of $d\rightarrow \tilde c_L$ production, there can be significant
competition between $R$-violating $\tilde c_L\rightarrow e^+d_R$ and
$R$-conserving $\tilde c_L\rightarrow c\chi$ decays, despite the small
$\lambda^\prime_{121}$ coupling, due to a possible cancellation in the
$R$-conserving coupling between the different gaugino components in the lightest
neutralino $\chi$~\cite{AEGLM}:
\beq
{1\over 2} g~(N_{i2} + {1\over 3} \tan\theta_W N_{i2})
\label{threeseventeen}
\eeq
The effect of this possible cancellation can be seen in Fig. 36 for $\mu > 0$.
On the other hand, in the case of $d\rightarrow \tilde t$ production, there is
only a very limited range of parameter space where  $\tilde t \rightarrow
e^+d$ is competitive with $R$-conserving $\tilde t$ decays~\cite{AEGLM}. The
case of
$s\rightarrow \tilde t$ production is intermediate between these two, in that
competition between the $R$-conserving and -violating decays is possible if
$m_{\tilde t} \gappeq$ 210 GeV, but more difficult for $m_{\tilde t} \lappeq$
200 GeV~\cite{ELS}. In any of these scenarios, $\tilde q\rightarrow q\chi$
decays could
provide an interesting alternative decay signature for either the HERA
experiments or CDF~\cite{AEGLM}.

\begin{figure}
\hglue2cm
\epsfig{figure=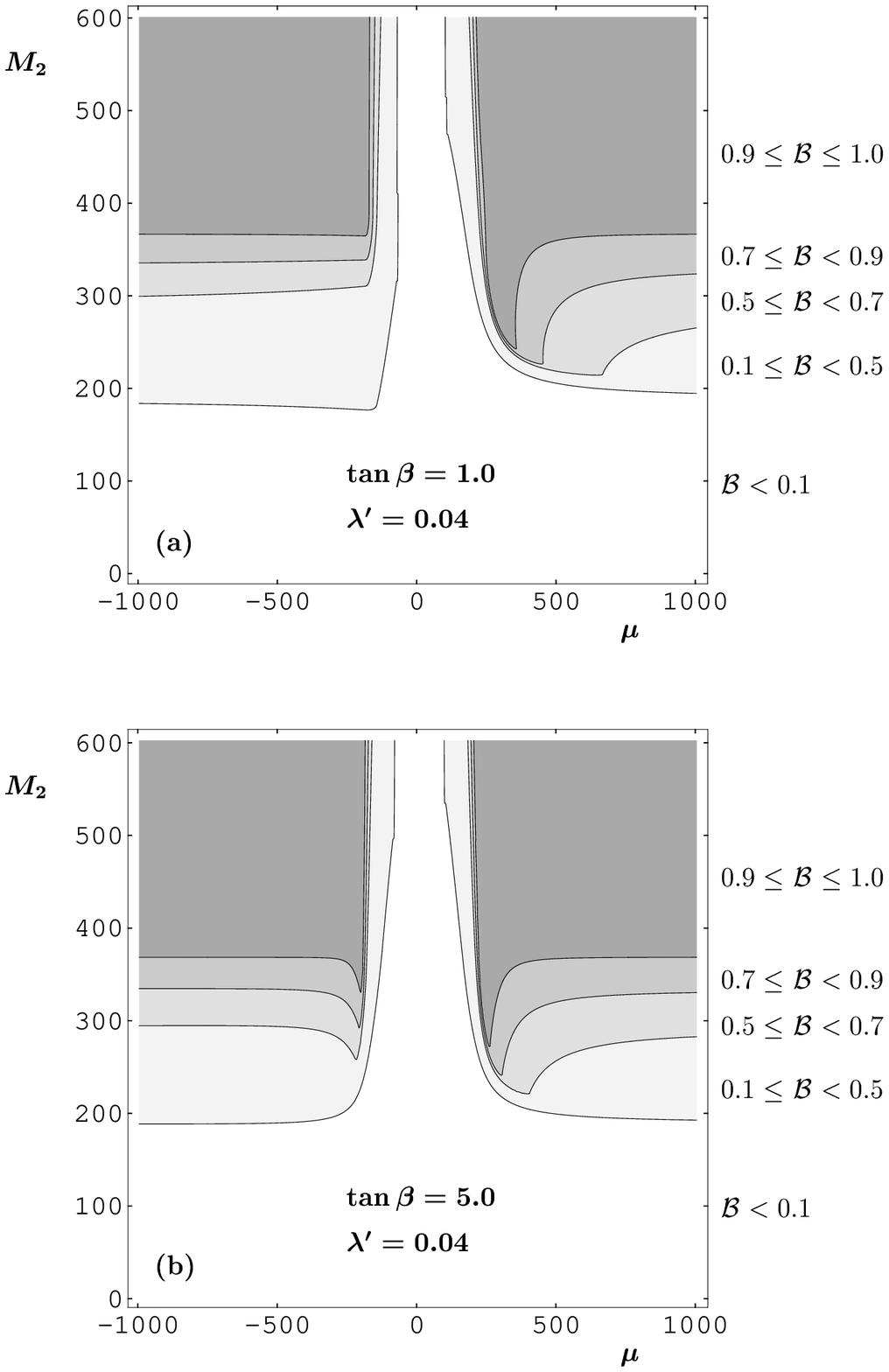,width=8cm}
\caption[]{
Branching ratio of $\tilde c_L\rightarrow c\chi$ decay in the $e^+d\rightarrow
\tilde c_L$ $R$-violating squark interpretation of the HERA large-$Q^2$ data.}
\end{figure}

What about signatures at LEP~2?  There could be effects on
$\sigma(e^+e^-\rightarrow \bar qq)$ and/or the final-state angular distributions
due to $R$-violating $\tilde q$ exchange~\cite{Zerwas}. There could be
single production
$e^+e^-\rightarrow e^\pm \stackrel{(-)}{\displaystyle q} \tilde q$ if $m_{\tilde
q} < E_{cm}$~\cite{single}. The reaction $e^+e^-\rightarrow \chi\chi$
would be detectable if
$m_\chi < E_{cm/2}$. Also interesting would be a $L_iL_jE^c_k$ coupling: this
could give interference effects or even a direct-channel $\tilde\nu$ resonance
in $e^+e^-\rightarrow L^+L^-$!~\cite{sneutrino}

\subsection{Supersymmetry at the LHC}

This has recently been the subject of a workshop at CERN~\cite{susywkshp},
and broad studies have
been made by the ATLAS and CMS collaborations using complementary approaches.
ATLAS has made detailed  analyses of a few selected points in the
multidimensional supersymmetric parameter space, and CMS has made a
comprehensive scan. A key feature of  these studies has been that the LHC
produces many heavier sparticles that decay into lighter ones, and that these
cascades can often be reconstructed efficiently~\cite{cascades}. In general,
the dominant
signatures are jets + leptons + $E\llap{$/$}_T$, and the physics reach
is large: $m_{\tilde q,\tilde g} \rightarrow$ 2 TeV,
$m_{\tilde \ell}\rightarrow$ 400 GeV, etc. Some examples of the reconstruction of
lighter sparticles in the cascade decays are shown in Fig. 37. These will
enable precision measurements of model parameters, within a given theoretical
framework~\cite{cascades}.

\begin{figure}
\hglue.2cm
\mbox{\epsfig{figure=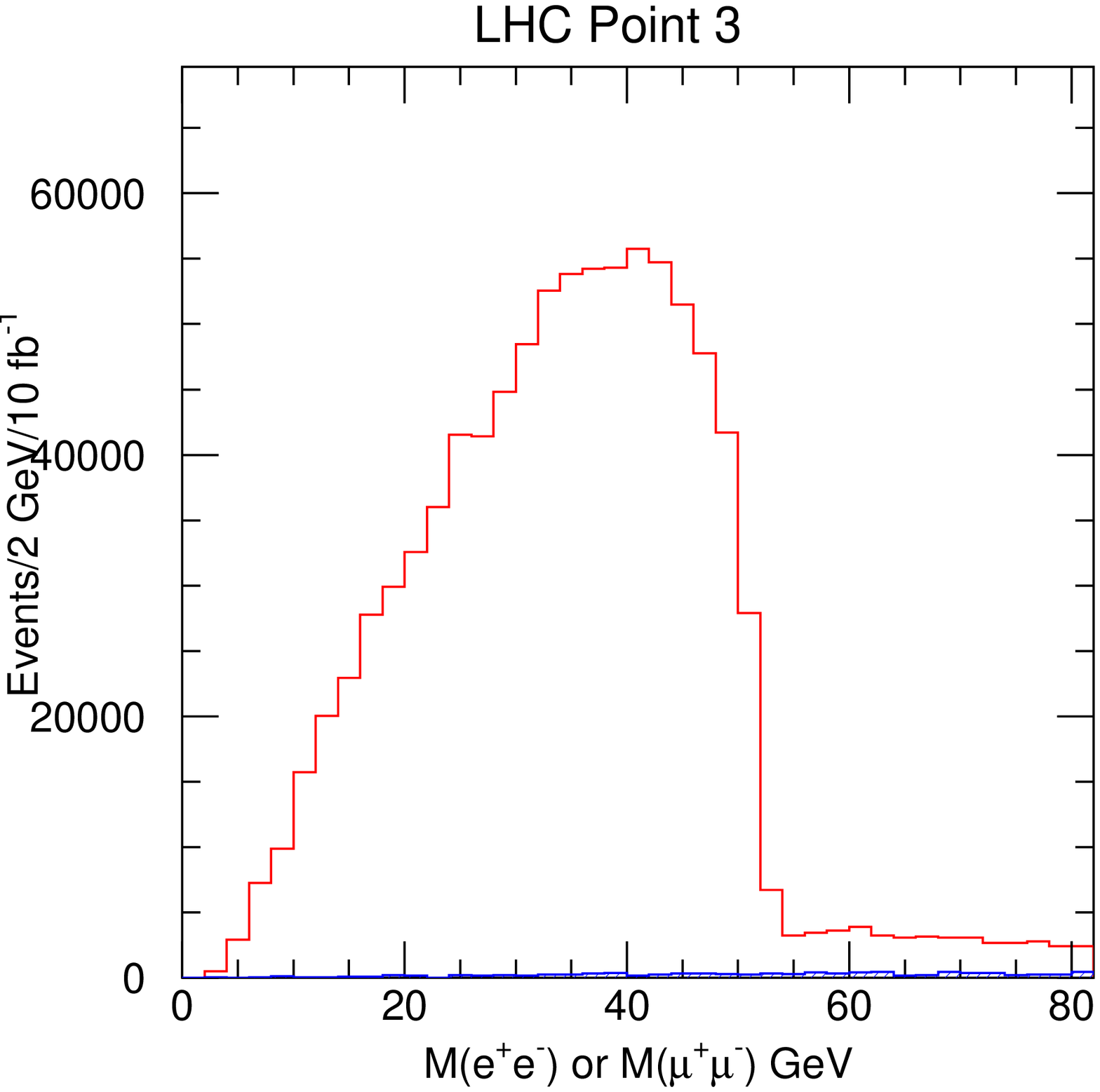,width=5cm}(a)
\epsfig{figure=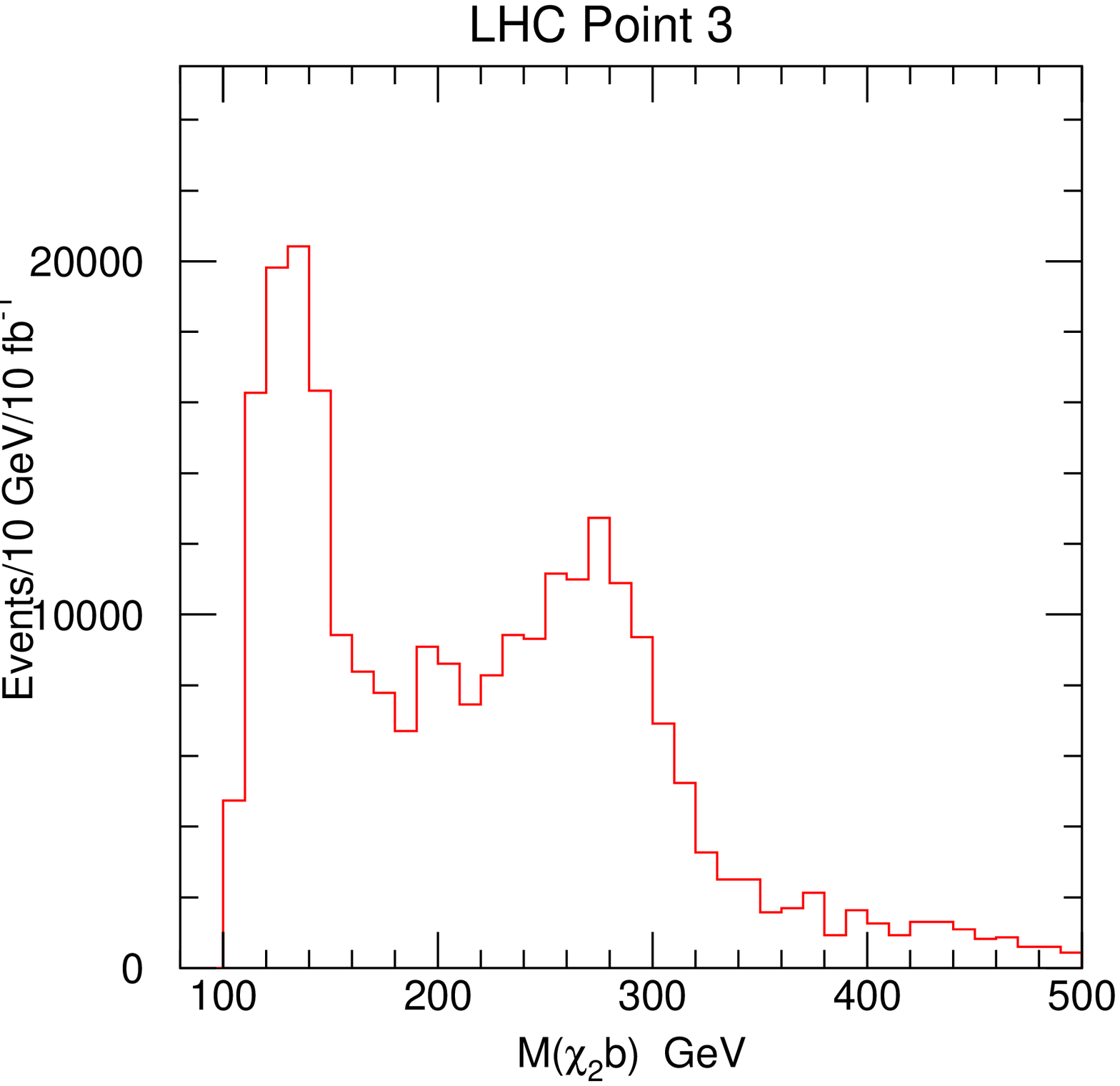,width=5cm}(b)}
\hglue.3cm
\begin{center}
\mbox{\epsfig{figure=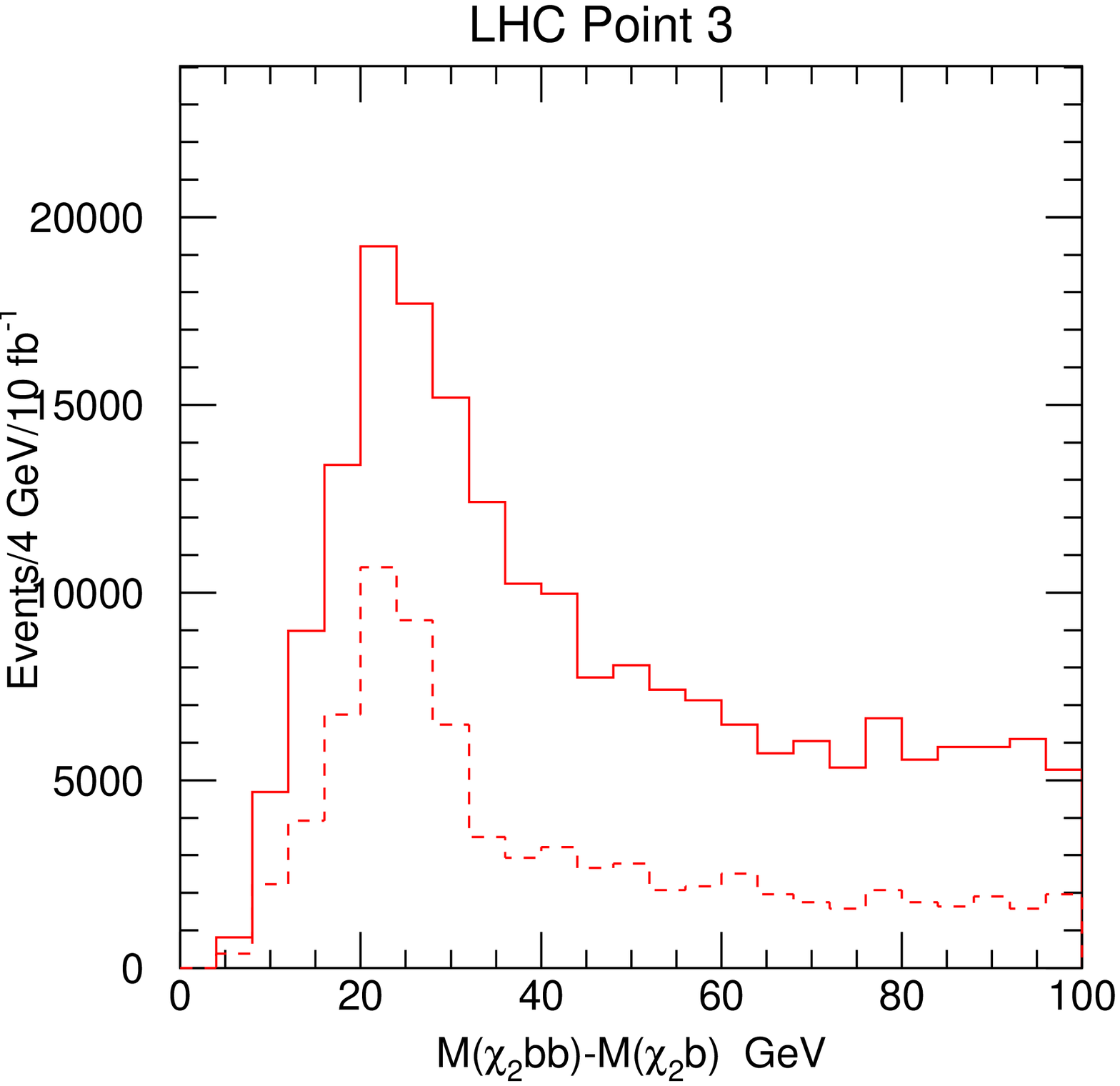,width=5cm}(c)}
\end{center}
\caption[]{
Reconstruction of sparticle masses for a particular supergravity model studied in
(cite{hcsusy}), demonstrating (a) the end-point in $\chi_2\rightarrow
\chi\ell^+\ell^-$ decay, (b) the $\tilde b\rightarrow \chi_2b$ mass bump, and (c)
the $\tilde g\rightarrow \tilde b\bar b$ mass bump.}
\end{figure}

Table 4 shows the plethora of sparticles detectable at each of the five points
in parameter space selected by ATLAS for special study. We see that the LHC may
be able to find a large fraction of the expected sparticle
Zooino~\cite{cascades}. Remember that
the Bevatron was built to find the antiproton (which it did), but is mainly
famous for discovering the complicated hadron spectrum. Much of the motivation
for the LHC is to find the Higgs boson (which it can), but maybe it will become
famous for finding sparticles -- the LHC as ``Bevatrino"?!

\begin{table}
\begin{center}
\caption{The LHC as ``Bevatrino":
Sparticles detectable at selectred points in supersymmetric
parameter space  are denoted by +}\vspace*{0.3cm}
\begin{tabular}{|c|cccccccccccc|}  \hline
 & $h$ & $H/A$& $\chi^0_2$ & $\chi^0_3$ & $\chi^0_4$ & $\chi^\pm_1$ &
$\chi^\pm_2$ & $\tilde q$ & $\tilde b$ & $\tilde t$ & $\tilde g$&
$\tilde\ell$ \\ \hline
1 & + && + & & & &  & + & + & + &+& \\
2 & + && + & & & &  & + & + & + &+& \\
3 & + &+& + &  && + && + & + && + & \\
4 & + && + & + & + & + & + & + &&&+ & \\
5 & + && + &&&&&+ & + & + & + & + \\
\hline
\end{tabular}
\end{center}
\end{table}

\end{document}